\pdfoutput=1 
\RequirePackage{luatex85}
\PassOptionsToPackage{dvipsnames}{xcolor}
\RequirePackage{amssymb}
\RequirePackage{varioref}
\PassOptionsToPackage{nameinlink}{cleverref}
\documentclass[a4paper,USenglish,cleveref,autoref,thm-restate,pdfa]{lipics-v2021}
\usepackage{hyperref}

\hideLIPIcs  

\nolinenumbers
\usepackage[frozencache,cachedir=_minted]{minted}
\usepackage{amsmath}
\usepackage{amssymb}
\usepackage{comment}
\usepackage[all]{xy}
\usepackage{stmaryrd}
\usepackage[utf8]{inputenc}
\usepackage{textgreek}
\usepackage{newunicodechar}
\usepackage{usebib}
\usepackage{tikz}
\usepackage{pgfplots}
\pgfplotsset{compat=1.18}
\usepackage{bcprules}
\usepackage{multicol}
\usepackage{apptools}
\usepackage{placeins}
\usepackage{adjustbox}
\usepackage{xparse}
\usepgfplotslibrary{units}
\usepgfplotslibrary{external}
\newunicodechar{→}{\ensuremath{\to}}
\newunicodechar{λ}{\ensuremath{\lambda}}
\newunicodechar{∀}{\ensuremath{\forall}}

\usepackage{etoolbox,xpatch}
\makeatletter
\AtBeginEnvironment{minted}{\dontdofcolorbox}
\def\dontdofcolorbox{\renewcommand\fcolorbox[4][]{##4}}
\xpatchcmd{\inputminted}{\minted@fvset}{\minted@fvset\dontdofcolorbox}{}{}
\xpatchcmd{\mintinline}{\minted@fvset}{\minted@fvset\dontdofcolorbox}{}{} 
\newcommand{\textmintinline}[2]{\text{\mintinline{#1}{#2}}}
\makeatother

\provideboolean{tobesplit}

\ifthenelse{\boolean{tobesplit}}{%
  \AtBeginDocument{}%
}{}
\makeatletter
\patchcmd{\hyper@makecurrent}{%
    \ifx\Hy@param\Hy@chapterstring
        \let\Hy@param\Hy@chapapp
    \fi
}{%
    \iftoggle{inappendix}{
        \@checkappendixparam{chapter}%
        \@checkappendixparam{section}%
        \@checkappendixparam{subsection}%
        \@checkappendixparam{subsubsection}%
        \@checkappendixparam{paragraph}%
        \@checkappendixparam{subparagraph}%
    }{}%
}{}{\show\hyper@makecurrent\errmessage{failed to patch}}

\newcommand*{\@checkappendixparam}[1]{%
    \def\@checkappendixparamtmp{#1}%
    \ifx\Hy@param\@checkappendixparamtmp
        \let\Hy@param\Hy@appendixstring
    \fi
}
\makeatletter

\newtoggle{inappendix}
\togglefalse{inappendix}

\apptocmd{\appendix}{\toggletrue{inappendix}}{}{\show\appendix\errmessage{failed to patch}}

\pgfplotsset{
    discard if not/.style 2 args={
        x filter/.code={
            \edef\tempa{\thisrow{#1}}
            \edef\tempb{#2}
            \ifx\tempa\tempb
            \else
                
            \fi
        }
    }
}

\newbibfield{author}
\newbibfield{title}
\newbibfield{citet}
\bibinput{rewriting-lower}

\newcommand{\citet}[1]{\usebibentry{#1}{citet}~\cite{#1}}

\newcommand{\coqbug}[1]{\href{https://github.com/coq/coq/issues/#1}{\##1}}
\newcommand{\coqpr}[1]{\href{https://github.com/coq/coq/pull/#1}{\##1}}

\NewDocumentCommand{\githublink}{m m o}{%
  \IfNoValueTF{#3}%
    {\href{https://github.com/#1/#2}{\texttt{#1/#2}}}%
    {\href{https://github.com/#1/#2/tree/#3}{\texttt{#1/#2@#3}}}%
}
\NewDocumentCommand{\githublinkurl}{m m o}{%
  \IfNoValueTF{#3}%
    {\url{https://github.com/#1/#2}}%
    {\url{https://github.com/#1/#2/tree/#3}}%
}
\newcommand{\Rtac}{\ensuremath{\mathcal{R}_\text{\emph{tac}}}}
\newcommand{\Ltac}{\ensuremath{\mathcal{L}_\text{\emph{tac}}}}
\newcommand{\tacvmcompute}{\mintinline{coq}{vm_compute}}
\newcommand{\tacnativecompute}{\mintinline{coq}{native_compute}}
\newcommand{\taccbv}{\mintinline{coq}{cbv}}

\newcommand{\tacsimpl}{\mintinline{coq}{simpl}}
\newcommand{\taccbn}{\textbf{\texttt{cbn}}}
\newcommand{\defeq}{=}
\makeatletter
\newcommand{\letin}[1][{\ensuremath{\cdots}}{\ensuremath{\cdots}}]{%
  \texttt{let }\@firstoftwo#1\texttt{ in }\@secondoftwo#1
}

\newcommand{\beginTikzpictureStamped}[2][]{%
  {%
    \everyeof{\noexpand}
    \long\xdef\@tikzstamp{#2}%
  }%
  \def\@dobegintikzpicture{\begin{tikzpicture}[#1]}%
  \expandafter\@dobegintikzpicture\expandafter\def\expandafter\tikzstamp\expandafter{\@tikzstamp}%
}
\newcommand{\TikzpictureStamped}[3][]{\beginTikzpictureStamped[#1]{#2}#3\end{tikzpicture}}%
\makeatother
\newcommand{\asserteq}[2]{\ifthenelse{\equal{#1}{#2}}{}{\GenericError{}{Not equal: \detokenize{#1} != \detokenize{#2}}{}{}}}

\makeatletter
\AtAppendix{%
  \newwrite\@appendixpagefile
  \immediate\openout\@appendixpagefile=\jobname.appendix.startpage\relax
  \immediate\write\@appendixpagefile{\detokenize{\def\appendixstartpage}{\thepage}}%
}
\newcommand{\startsupplement}{%
  \clearpage
  \newwrite\@supplementpagefile
  \immediate\openout\@supplementpagefile=\jobname.supplement.startpage\relax
  \immediate\write\@supplementpagefile{\detokenize{\def\supplementstartpage}{\thepage}}%
  \newwrite\@supplementonlypagefile
  \immediate\openout\@supplementonlypagefile=\jobname.supplement.onlystartpage\relax
  \immediate\write\@supplementonlypagefile{\thepage}%
}%
\AtEndDocument{%
  \newwrite\@endsupplementonlypagefile
  \immediate\openout\@endsupplementonlypagefile=\jobname.supplement.onlyendpage\relax
  \immediate\write\@endsupplementonlypagefile{\thepage}%
}
\makeatother

\title{Accelerating Verified-Compiler Development with a Verified Rewriting Engine}

\author{Jason Gross}{CSAIL, Massachusetts Institute of Technology, Cambridge, MA, USA \and Machine Intelligence Research Institute, Berkeley, CA, USA \and \url{https://jasongross.github.io/} }{jgross@mit.edu}{https://orcid.org/0000-0002-9427-4891}{}

\author{Andres Erbsen}{CSAIL, Massachusetts Institute of Technology, Cambridge, MA, USA \and \url{https://andres.systems/} }{andreser@mit.edu}{}{}

\author{Jade Philipoom
}{CSAIL, Massachusetts Institute of Technology, Cambridge, MA, USA \and Google, London, UK}{jade.philipoom@gmail.com}{}{}

\author{Rajashree Agrawal}{Reed College, Portland, OR, USA
}{ragrawal@reed.edu}{https://orcid.org/0000-0001-7617-9180}{}

\author{Adam Chlipala}{CSAIL, Massachusetts Institute of Technology, Cambridge, MA, USA \and \url{http://adam.chlipala.net/} }{adamc@csail.mit.edu}{https://orcid.org/0000-0001-7085-9417}{
}

\authorrunning{J. Gross, A. Erbsen, J. Philipoom, R. Agrawal, and A. Chlipala} 

\Copyright{Jason Gross, Andres Erbsen, Jade Philipoom, Rajashree Agrawal, and Adam Chlipala} 



\ifthenelse{\boolean{tobesplit}}{%
}{}

\supplement{ }
\ifthenelse{\boolean{tobesplit}}{%
}{}
\supplementdetails[
  subcategory={Source Code}, swhid={swh:1:rev:1787ab401a7e71afc9937010e2e155e4b1594ab5;origin=https://github.com/mit-plv/rewriter;visit=swh:1:snp:6d296391d575112bf64de359f5ee0efad7b89305}
]{Software}{https://github.com/mit-plv/rewriter/tree/ITP-2022-perf-data} 
\supplementdetails[
  subcategory={Source Code}, swhid={swh:1:rev:72fe0dddee5e6dceeab0b8a2e6a745abf5287d3e;origin=https://github.com/mit-plv/fiat-crypto;visit=swh:1:snp:6d8205c7c35b7b6fd767dbebd2089f4384ca4691}
]{Software}{https://github.com/mit-plv/fiat-crypto/tree/perf-testing-data-ITP-2022-rewriting} 

\funding{%
  This work was supported in part by a Google Research Award, National Science Foundation grants CCF-1253229, CCF-1512611, and CCF-1521584, and the National Science Foundation Graduate Research Fellowship under Grant Nos.\ 1122374 and 1745302.
  Any opinion, findings, and conclusions or recommendations expressed in this material are those of the authors and do not necessarily reflect the views of the National Science Foundation.%
}

\EventEditors{June Andronick and Leonardo de Moura}
\EventNoEds{2}
\EventLongTitle{13th International Conference on Interactive Theorem Proving (ITP 2022)}
\EventShortTitle{ITP 2022}
\EventAcronym{ITP}
\EventYear{2022}
\EventDate{August 7--10, 2022}
\EventLocation{Haifa, Israel}
\EventLogo{}
\SeriesVolume{237}
\ArticleNo{5}

\begin{document}

\maketitle

\begin{abstract}
  Compilers are a prime target for formal verification, since compiler bugs invalidate higher-level correctness guarantees, but compiler changes may become more labor-intensive to implement, if they must come with proof patches.
  One appealing approach is to present compilers as sets of algebraic rewrite rules, which a generic engine can apply efficiently.
  Now each rewrite rule can be proved separately, with no need to revisit past proofs for other parts of the compiler.
  We present the first realization of this idea, in the form of a framework for the Coq proof assistant.
  Our new Coq command takes normal proved theorems and combines them automatically into fast compilers with proofs.
  We applied our framework to improve the Fiat Cryptography toolchain for generating cryptographic arithmetic, producing an extracted command-line compiler that is about 1000$\times$ faster while actually featuring simpler compiler-specific proofs.
\end{abstract}

\section{Introduction}\label{sec:intro}

Formally verified compilers like CompCert~\cite{Compcert} and CakeML~\cite{CakeML} are success stories for proof assistants, helping close a trust gap for one of the most important categories of software infrastructure.
A popular compiler cannot afford to stay still; developers will add new backends, new language features, and better optimizations.
Proofs must be adjusted as these improvements arrive.
It makes sense that the author of a new piece of compiler code must prove its correctness, but ideally there would be no need to revisit old proofs.
There has been limited work, though, on avoiding that kind of coupling.
\citet{xcert} demonstrated a streamlined way to extend CompCert with new verified optimizations driven by dataflow analysis, but we are not aware of past work that supports easy extension for compilers from functional languages to C code.
We present our work targeting that style.

One strategy for writing compilers modularly is to exercise foresight in designing a core that will change very rarely, such that feature iteration happens outside the core.
Specifically, phrasing the compiler in terms of rewrite rules allows clean abstractions and conceptual boundaries~\cite{Hickey2006}.
Then, most desired iteration on the compiler can be achieved through iteration on the rewrite rules.

It is surprisingly difficult to realize this modular approach with good performance.
Verified compilers can either be proof-producing (certifying) or proven-correct (certified).
Proof-producing compilers usually operate on the functional languages of the proof assistants that they are written in, and variable assignments are encoded as let binders.
All existing proof-producing rewriting strategies scale at least quadratically in the number of binders.
This performance scaling is inadequate for applications like Fiat Cryptography~\cite{FiatCryptoSP19} where the generated code has 1000s of variables in a single function.
Proven-correct compilers do not suffer from this asymptotic blowup in the number of binders.

In this paper, we present \textbf{the first proven-correct compiler-builder toolkit parameterized on rewrite rules}.
Arbitrary sets of Coq theorems (quantified equalities) can be assembled by a single new Coq command into an extraction-ready verified compiler.
We did not need to extend the trusted code base, so our compiler compiler need not be trusted.
We achieve both good performance of compiler runs and good performance of generated code, via addressing a number of scale-up challenges vs.\ past work.

We evaluate our toolkit by replacing a key component of Fiat Cryptography~\cite{FiatCryptoSP19}, a Coq library that generates code for big-integer modular arithmetic at the heart of elliptic-curve-cryptography algorithms.
Routines generated (with proof) with Fiat Cryptography now ship with all major Web browsers and all major mobile operating systems.
With our improved compiler architecture, it became easy to add two new backends and a variety of new supported source-code features, and we were easily able to try out new optimizations.

Replacing Fiat Cryptography's original compiler with the compiler generated by our toolkit has two additional benefits.
Fiat Cryptography was previously only used successfully to build C code for two of the three most widely used curves (P-256 and Curve25519).
Our prior version's execution timed out trying to compile code for the third most widely used curve (P-384).
Using our new toolkit has made it possible to generate compiler-synthesized code for P-384 while generating completely identical code for the primes handled by the previous version, about 1000$\times$ more quickly.
Additionally, Fiat Cryptography previously required source code to be written in continuation-passing style, and our compiler has enabled a direct-style approach, which pays off in simplifying theorem statements and proofs.

\subsection{Related Work}\label{sec:related-one}

Assume our mission is to take libraries of purely functional combinators, apply them to compile-time parameters, and compile the results down to lean C code.
Furthermore, we ask for machine-checked proofs that the C programs preserve the behavior of the higher-order functional programs we started with.
What good ideas from the literature can we build on?

\citet{Hickey2006} discuss at length how to build compilers around rewrite rules.
``All program transformations, from parsing to code generation, are cleanly isolated and specified as term rewrites.''
While they note that the correctness of the compiler is thus reduced to the correctness of the rewrite rules, they did not prove correctness mechanically.
Furthermore, it is not clear that they manage to avoid the asymptotic blow-up associated with proof-producing rewriting of deeply nested let-binders.
They give no performance numbers, so it is hard to say whether or not their compiler performs at the scale necessary for Fiat Cryptography.
Their rewrite-engine driver is unproven OCaml code, while we will produce custom drivers with Coq proofs.

\Rtac{}~\cite{rtac} is a more general framework for verified proof tactics in Coq, including an experimental reflective version of \texttt{rewrite\_strat} supporting arbitrary setoid relations, unification variables, and arbitrary semidecidable side conditions solvable by other verified tactics, using de Bruijn indexing to manage binders.
We found that \Rtac{} misses a critical feature for compiling large programs: preserving subterm sharing.
As a result, our experiments with compiler construction yielded clear asymptotic slowdown vs.\ what we eventually accomplished.
\Rtac{} is also more heavyweight to use, for instance requiring that theorems be restated manually in a deep embedding to bring them into automation procedures.
Furthermore, we are not aware of any past experiments driving verified compilers with \Rtac{}.

\citet{Aehlig} came closest to a fitting approach, using \emph{normalization by evaluation (NbE)}~\cite{NbE} to bootstrap reduction of open terms on top of full reduction, as built into a proof assistant.
However, it was simultaneously true that they expanded the proof-assistant trusted code base in ways specific to their technique, and that they did not report any experiments actually using the tool for partial evaluation (just traditional full reduction), potentially hiding performance-scaling challenges or other practical issues.
For instance, they also do not preserve subterm sharing explicitly, and they represent variable references as unary natural numbers (de Bruijn-style).
They also require that rewrite rules be embodied in ML code, rather than stated as natural ``native'' lemmas of the proof assistant.
We will follow their basic outline with important modifications.

Our implementation builds on fast full reduction in Coq's kernel, via a virtual machine~\cite{vmcompute} or compilation to native code~\cite{nativecompute} (neither verified).
Especially the latter is similar in adopting NbE for full reduction, simplifying even under $\lambda$s, on top of a more traditional implementation of OCaml that never executes preemptively under $\lambda$s.
Neither approach unifies support for rewriting with proved rules, and partial evaluation only applies in very limited cases.

A variety of forms of pragmatic partial evaluation have been demonstrated, with Lightweight Modular Staging~\cite{LMS} in Scala as one of the best-known current examples.
The LMS-Verify system~\cite{LMSVerify} can be used for formal verification of generated code after-the-fact.
Typically LMS-Verify has been used with relatively shallow properties (though potentially applied to larger and more sophisticated code bases than we tackle), not scaling to the kinds of functional-correctness properties that concern us here.

So, overall, to our knowledge, no past compiler as a set of rewrite rules has come with a full proof of correctness as a standalone functional program.
Related prior work with mechanized proofs suffered from both performance bottlenecks and usability problems, the latter in requiring that eligible rewrite rules be stated in special deep embeddings.

\subsection{Our Solution}\label{sec:our-solution}

Our variant on the technique of \citet{Aehlig} has these advantages:

\begin{itemize}
\item It integrates with a general-purpose, foundational proof assistant, \textbf{without growing the trusted code base}.
\item For a wide variety of initial functional programs, it provides \textbf{fast} partial evaluation with reasonable memory use.
\item It allows reduction that \textbf{mixes} \emph{rules of the definitional equality} with \emph{equalities proven explicitly as theorems}.
\item It allows \textbf{rapid iteration} on rewrite rules with \emph{minimal verification overhead}.
\item It \textbf{preserves sharing} of common subterms.
\item It also allows \textbf{extraction of standalone compilers}.
\end{itemize}

Our contributions include answers to a number of challenges that arise in scaling NbE-based partial evaluation in a proof assistant.
First, we rework the approach of \citet{Aehlig} to function \emph{without extending a proof assistant's trusted code base}, which, among other challenges, requires us to prove termination of reduction and encode pattern matching explicitly (leading us to adopt the performance-tuned approach of \citet{maranget2008compiling}).
We also improve on Coq-specific related work (e.g., of \citet{rtac}) by allowing rewrites to be written in natural Coq form (not special embedded syntax-tree types), while supporting optimizations associated with past unverified engines (e.g., \citet{Boespflug2009}).

Second, using partial evaluation to generate residual terms thousands of lines long raises \emph{new scaling challenges}:
\begin{itemize}
\item
  Output terms may contain so \emph{many nested variable binders} that we expect it to be performance-prohibitive to perform bookkeeping operations on first-order-encoded terms (e.g., with de Bruijn indices, as is done in \Rtac{} by \citet{rtac}).
  For instance, while the reported performance experiments of \citet{Aehlig} generate only closed terms with no binders, Fiat Cryptography may generate a single routine (e.g., multiplication for curve P-384) with nearly a thousand nested binders.
\item
  Naive representation of terms without proper \emph{sharing of common subterms} can lead to fatal term-size blow-up.
\item
  Unconditional rewrite rules are in general insufficient, and we need \emph{rules with side conditions}.
  E.g., Fiat Cryptography depends on checking lack-of-overflow conditions.
\item
  However, it is also not reasonable to expect a general engine to discharge all side conditions on the spot.
  We need integration with \emph{abstract interpretation}.
\end{itemize}

Briefly, our respective solutions to these problems are the \emph{parametric higher-order abstract syntax (PHOAS)}~\cite{PhoasICFP08} term encoding, a \emph{let-lifting} transformation threaded throughout reduction, extension of rewrite rules with executable Boolean side conditions, and a design pattern that uses decorator function calls to include analysis results in a program.

Finally, we carry out the \emph{first large-scale performance-scaling evaluation} of a verified rewrite-rule-based compiler, covering all elliptic curves from the published Fiat Cryptography experiments, along with microbenchmarks.

We pause to give a motivating example before presenting the core structure of our engine (\autoref{sec:structure}), the additional scaling challenges we faced (\autoref{sec:scaling}), experiments (\autoref{sec:evaluation}), and conclusions.
Our implementation is attached.

\section{A Motivating Example}\label{sec:motivating-example}\label{sec:explain-ident.eagerly}\label{sec:explain-eval-rect}

Our compilation style involves source programs that mix higher-order functions and inductive types.
We want to compile to C code, reducing away uses of fancier features while seizing opportunities for arithmetic simplification.
Here is a small but illustrative example.

\begin{minted}[fontsize=\small]{coq}
Definition prefixSums (ls : list nat) : list nat :=
  let ls' := combine ls (seq 0 (length ls)) in
  let ls'' := map (λ p, fst p * snd p) ls' in
  let '(_, ls''') := fold_left (λ '(acc, ls''') n,
    let acc' := acc + n in (acc', acc' :: ls''')) ls'' (0, []) in ls'''.
\end{minted}

This function first computes list \mintinline{coq}{ls'} that pairs each element of input list \mintinline{coq}{ls} with its position, so, for instance, list $[a; b; c]$ becomes $[(a, 0); (b, 1); (c, 2)]$.
Then we map over the list of pairs, multiplying the components at each position.
Finally, we compute all prefix sums.

We would like to specialize this function to particular list lengths.
That is, we know in advance how many list elements we will pass in, but we do not know the values of those elements.
For a given length, we can construct a schematic list with one free variable per element.
For example, to specialize to length four, we can apply the function to list \mintinline{coq}{[a; b; c; d]}, and we expect this output:
\begin{minted}[fontsize=\small]{coq}
let acc := b + c * 2 in let acc' := acc + d * 3 in [acc'; acc; b; 0]
\end{minted}

We do not quite have C code yet, but, composing this code with another routine to consume the output list, we easily arrive at a form that looks almost like three-address code and is quite easy to translate to C and many other languages.

Notice how subterm sharing via \mintinline{coq}{let}s is important.
As list length grows, we avoid quadratic blowup in term size through sharing.
Also notice how we simplified the first two multiplications with $a \cdot 0 = 0$ and $b \cdot 1 = b$ (each of which requires explicit proof in Coq), using other arithmetic identities to avoid introducing new variables for the first two prefix sums of \mintinline{coq}{ls''}, as they are themselves constants or variables, after simplification.

To set up our compiler, we prove the algebraic laws that it should use for simplification, starting with basic arithmetic identities.
\begin{minted}[fontsize=\small]{coq}
Lemma zero_plus : ∀ n, 0 + n = n.      Lemma times_zero : ∀ n, n * 0 = 0.
Lemma plus_zero : ∀ n, n + 0 = n.      Lemma times_one  : ∀ n, n * 1 = n.
\end{minted}

Next, we prove a law for each list-related function, connecting it to the primitive-recursion combinator for some inductive type (natural numbers or lists, as appropriate).
We also use a further marker \mintinline{coq}{ident.eagerly} to ask the compiler to simplify a case of primitive recursion by complete traversal of the designated argument's constructor tree.
\begin{minted}[fontsize=\small]{coq}
Lemma eval_map A B (f : A -> B) l
: map f l = ident.eagerly list_rect _ _ [] (λ x _ l', f x :: l') l.
Lemma eval_fold_left A B (f : A -> B -> A) l a
: fold_left f l a = ident.eagerly list_rect _ _ (λ a, a) (λ x _ r a, r (f a x)) l a.
Lemma eval_combine A B (la : list A) (lb : list B)
: combine la lb =
list_rect _ (λ _, []) (λ x _ r lb, list_case (λ _, _) [] (λ y ys, (x,y)::r ys) lb) la lb.
Lemma eval_length A (ls : list A)
: length ls = list_rect _ 0 (λ _ _ n, S n) ls.
\end{minted}

With all the lemmas available, we can package them up into a rewriter, which triggers generation of a specialized compiler and its soundness proof.
Our Coq plugin introduces a new command \mintinline{coq}{Make} for building rewriters
\begin{minted}[fontsize=\small]{coq}
Make rewriter := Rewriter For (zero_plus, plus_zero, times_zero, times_one, eval_map,
  eval_fold_left, do_again eval_length, do_again eval_combine,
  eval_rect nat, eval_rect list, eval_rect prod) (with delta) (with extra idents (seq)).
\end{minted}
Most inputs to \mintinline{coq}{Rewriter For} list quantified equalities to use for left-to-right rewriting.
However, we also use options \mintinline{coq}{do_again}, to request that some rules trigger extra bottom-up passes after being used for rewriting; \mintinline{coq}{eval_rect}, to queue up eager evaluation of a call to a primitive-recursion combinator on a known recursive argument; \mintinline{coq}{with delta}, to request evaluation of all monomorphic operations on concrete inputs; and \mintinline{coq}{with extra idents}, to inform the engine of further permitted identifiers that do not appear directly in any of the rewrite rules.

Our plugin also provides new tactics like \mintinline{coq}{Rewrite_rhs_for}, which applies a rewriter to the right-hand side of an equality goal.
That last tactic is just what we need to synthesize a specialized \mintinline{coq}{prefixSums} for list length four, along with its correctness proof.
\begin{minted}[fontsize=\small]{coq}
Definition prefixSums4 :
{f:nat→nat→nat→nat→list nat | ∀ a b c d, f a b c d = prefixSums [a;b;c;d]}
  := ltac:(eexists; Rewrite_rhs_for rewriter; reflexivity).
\end{minted}

That compiler execution ran inside of Coq, but an even more pragmatic approach is to \emph{extract} the compiler as a standalone program in OCaml or Haskell.
Such a translation is possible because the \mintinline{coq}{Make} command produces a proved program in Gallina, Coq's logic.
As a result, our reworking of Fiat Cryptography compilation culminated in extraction of a command-line OCaml program that developers in industry have been able to run without our help, where Fiat Cryptography previously required installing and running Coq, with an elaborate build process to capture its output.
It is also true that the standalone program is about 10$\times$ as fast as execution within Coq, though the trusted code base is larger.


\section{The Structure of a Rewriter}\label{sec:structure}
We are mostly guided by \citet{Aehlig} but made a number of crucial changes.
Let us review the basic idea of the approach of Aehlig et al.
First, their supporting library contains:
\begin{enumerate}
\item
  Within the logic of the proof assistant (Isabelle/HOL, in their case), a type of syntax trees for ML programs is defined, with an associated (trusted) operational semantics.
\item
  They also wrote a reduction function in (deeply embedded) ML, parameterized on a function to choose the next rewrite, and proved it sound once-and-for-all.
\end{enumerate}

Given a set of rewrite rules and a term to simplify, their main tactic must:
\begin{enumerate}
\item
  \emph{Generate a (deeply embedded) ML program that decides which rewrite rule, if any, to apply at the top node of a syntax tree}, along with a proof of its soundness.
\item
  \emph{Generate a (deeply embedded) ML term standing for the term we set out to simplify}, with a proof that it means the same as the original.
\item
  Combining the general proof of the rewrite engine with proofs generated by reification (the prior two steps), conclude that an application of the reduction function to the reified rules and term is indeed an ML term that generates correct answers.
\item
  ``Throw the ML term over the wall,'' using a general code-generation framework for Isabelle/HOL~\cite{CodeGen}.
  Trusted code compiles the ML code into the concrete syntax of Standard ML, and compiles it, and runs it, asserting an axiom about the outcome.
\end{enumerate}

Here is where our approach differs at that level of detail:
\begin{itemize}
\item
  Our reduction engine is written \emph{as a normal Gallina functional program}, rather than within a deeply embedded language.
  As a result, we are able to prove its type-correctness and termination, and we are able to run it within Coq's kernel.
\item
  We do \emph{compile-time specialization of the reduction engine} to sets of rewrite rules, removing overheads of generality.
\end{itemize}

\subsection{Our Approach in Ten Steps}\label{sec:nine-steps}

Here is a bit more detail on the steps that go into applying our Coq plugin, many of which we expand on in the following sections.
For \mintinline{coq}{Make} to precompute a rewriter:
\begin{enumerate}
\item
  The given lemma statements are scraped for which named identifiers to encode.
\item
  Inductive types enumerating all available primitive types and functions are emitted.
  This allows us to achieve the performance gains attributed in \citet{Boespflug2009} to having native metalanguage constructors for all constants, without manual coding.
\item
  Tactics generate all of the necessary definitions and prove all of the necessary lemmas for dealing with this particular set of inductive codes.
  Definitions include operations like Boolean equality on type codes and lemmas like ``all types have decidable equality.''
\item
  The statements of rewrite rules are reified and soundness and syntactic-well-formedness lemmas are proven about each of them.
\item
  Definitions and lemmas needed to prove correctness are assembled into a single package.
\end{enumerate}

Then, to rewrite in a goal, the following steps are performed:
\begin{enumerate}
\item
  Rearrange the goal into a single quantifier-free logical formula.
\item
  Reify a selected subterm and replace it with a call to our denotation function.
\item
  Rewrite with a theorem, into a form calling our rewriter.
\item Call Coq's built-in full reduction (\tacvmcompute{}) to reduce this application.
\item
  Run standard call-by-value reduction to simplify away use of the denotation function. 
\end{enumerate}


The object language of our rewriter is nearly simply typed. 
\begin{align*}
  e ::={}& \phantom{\mid} \texttt{App }e_1\texttt{ }e_2 \mid \texttt{Let }v \defeq e_1\texttt{ In }e_2 
  \mid \texttt{Abs }(\lambda v.\,e) \mid \texttt{Var }v \mid \texttt{Ident }i
\end{align*}
The \texttt{Ident} case is for identifiers, which are described by an enumeration specific to a use of our library.
For example, the identifiers might be codes for $+$, $\cdot$, and literal constants.
We write $\llbracket e \rrbracket$ for a standard denotational semantics.\label{sec:denote-brackets-def}

\subsection{Pattern-Matching Compilation and Evaluation}\label{sec:pattern-matching-compilation-and-evaluation}

\citet{Aehlig} feed a specific set of user-provided rewrite rules to their engine by generating code for an ML function, which takes in deeply embedded term syntax (actually \emph{doubly} deeply embedded, within the syntax of the deeply embedded ML!) and uses ML pattern matching to decide which rule to apply at the top level.
Thus, they delegate efficient implementation of pattern matching to the underlying ML implementation.
As we instead build our rewriter in Coq's logic, we have no such option to defer to ML.

We could follow a naive strategy of repeatedly matching each subterm against a pattern for every rewrite rule, as in the rewriter of \citet{rtac}, but in that case we do a lot of duplicate work when rewrite rules use overlapping function symbols.
Instead, we adopted the approach of \citet{maranget2008compiling}, who describes compilation of pattern matches in OCaml to decision trees that eliminate needless repeated work (for example, decomposing an expression into $x + y + z$ only once even if two different rules match on that pattern).

There are three steps to turn a set of rewrite rules into a functional program that takes in an expression and reduces according to the rules.
The first step is pattern-matching compilation: we must compile the left-hand sides of the rewrite rules to a decision tree that describes how and in what order to decompose the expression, as well as describing which rewrite rules to try at which steps of decomposition.
Because the decision tree is merely a decomposition hint, we require no proofs about it to ensure soundness of our rewriter.
The second step is decision-tree evaluation, during which we decompose the expression as per the decision tree, selecting which rewrite rules to attempt.
The only correctness lemma needed for this stage is that any result it returns is equivalent to picking some rewrite rule and rewriting with it.
The third and final step is to actually rewrite with the chosen rule.
Here the correctness condition is that we must not change the semantics of the expression.

While pattern matching begins with comparing one pattern against one expression, Maranget's approach works with intermediate goals that check multiple patterns against multiple expressions.
A decision tree describes how to match a vector (or list) of patterns against a vector of expressions.
It is built from these constructors:
\begin{itemize}
  \item \texttt{TryLeaf k onfailure}: Try the $k^\text{th}$ rewrite rule; if it fails, keep going with \texttt{onfailure}.
  \item \texttt{Failure}: Abort; nothing left to try.
  \item \texttt{Switch icases app\_case default}:
    With the first element of the vector, match on its kind; if it is an identifier matching something in \texttt{icases}, which is a list of pairs of identifiers and decision trees, remove the first element of the vector and run that decision tree; if it is an application and \texttt{app\_case} is not \texttt{None}, try the \texttt{app\_case} decision tree, replacing the first element of each vector with the two elements of the function and the argument it is applied to; otherwise, do not modify the vectors and use the \texttt{default}.
  \item \texttt{Swap i cont}: Swap the first element of the vector with the $i^\texttt{th}$ element (0-indexed) and keep going with \texttt{cont}.
\end{itemize}

Consider the encoding of two simple example rewrite rules, where we follow Coq's \Ltac{} language in prefacing pattern variables with question marks.
\begin{align*}
  ?n + 0 & \to n 
  &
  \texttt{fst}_{\mathbb{Z},\mathbb{Z}}(?x, ?y) & \to x
\end{align*}
We embed them in an AST type for patterns, which largely follows our ASTs for expressions.
\begin{verbatim}
0. App (App (Ident +) Wildcard) (Ident (Literal 0))
1. App (Ident fst) (App (App (Ident pair) Wildcard) Wildcard)
\end{verbatim}
The decision tree produced is\label{sec:compiled-pattern}
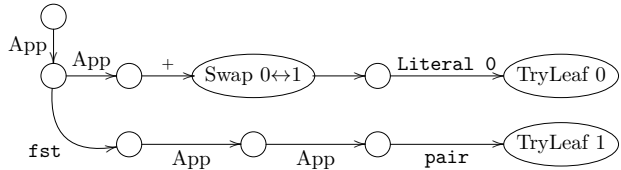
\[\resizebox{230px}{!}{\xymatrix@R-1pc{
  *++[o][F-]\txt{} \ar[d]_{\txt{App}} \\
  *++[o][F-]\txt{} \ar[r]^-{\txt{App}} \ar@/_1.5pc/[dr]_{\txt{\texttt{fst}}} & *++[o][F-]\txt{} \ar[r]^-{+} & *++[o][F-]\txt{Swap 0$\leftrightarrow$1} \ar[r] & *++[o][F-]\txt{} \ar[rr]^-{\txt{\texttt{Literal~0}}} && *++[o][F-]\txt{TryLeaf 0} \\
  & *++[o][F-]\txt{} \ar[r]_-{\txt{App}} & *++[o][F-]\txt{} \ar[r]_-{\txt{App}} & *++[o][F-]\txt{} \ar[rr]_-{\txt{\texttt{pair}}} && *++[o][F-]\txt{TryLeaf 1}
}}\]
\noindent where every nonswap node implicitly has a ``default'' case arrow to \texttt{Failure} and circles represent \texttt{Switch} nodes.

We implement, in Coq's logic, an evaluator for these trees against terms.
Note that we use Coq's normal partial evaluation to turn our general decision-tree evaluator into a specialized matcher to get reasonable efficiency.
Although this partial evaluation of our partial evaluator is subject to the same performance challenges we highlighted in the introduction, it only has to be done once for each set of rewrite rules, and we are targeting cases where the time of per-goal reduction dominates this time of metacompilation.

For our running example of two rules, specializing gives us this match expression.
\begin{minted}[fontsize=\small]{coq}
match e with
| App f y => match f with
  | Ident fst => match y with
    | App (App (Ident pair) x) y => x | _ => e end
  | App (Ident +) x => match y with
    | Ident (Literal 0) => x | _ => e end | _ => e end | _ => e end.
\end{minted}

\subsection{Adding Higher-Order Features}\label{sec:thunk-eval-subst-term}

Fast rewriting at the top level of a term is the key ingredient for supporting customized algebraic simplification.
However, not only do we want to rewrite throughout the structure of a term, but we also want to integrate with simplification of higher-order terms, in a way where we can prove to Coq that our syntax-simplification function always terminates.
Normalization by evaluation (NbE)~\cite{NbE} is an elegant technique for adding the latter aspect, in a way where we avoid needing to implement our own $\lambda$-term reducer or prove it terminating.

To orient expectations: we would like to enable the following reduction
\begin{align*}
  (\lambda f\ x\ y.\, f\ x\ y)\ (+)\ z\ 0 & \leadsto z
\end{align*}
\noindent using the rewrite rule
\begin{align*}
  ?n + 0 & \to n
\end{align*}

We begin by reviewing NbE's most classic variant, for performing full $\beta$-reduction in a simply typed term in a guaranteed-terminating way.
Our simply typed $\lambda$-calculus syntax is:
\begin{align*}
  t & ::= t \to t ~|~ b
  & e & ::= \lambda v.\, e ~|~ e~e ~|~ v ~|~ c
\end{align*}
\noindent with $v$ for variables, $c$ for constants, and $b$ for base types.

We can now define normalization by evaluation.
First, we choose a ``semantic'' representation for each syntactic type, which serves as an interpreter's result type.
\begin{align*}
  \text{NbE}_t(t_1 \to t_2) & \defeq \text{NbE}_t(t_1) \to \text{NbE}_t(t_2)
  & \text{NbE}_t(b) & \defeq \texttt{expr}(b)
\end{align*}
Function types are handled as in a simple denotational semantics, while base types receive the perhaps-counterintuitive treatment that the result of ``executing'' one is a syntactic expression of the same type.
We write $\texttt{expr}(b)$ for the metalanguage type of object-language syntax trees of type $b$, relying on a type family $\texttt{expr}$.

\begin{figure}
  \begin{subfigure}[b]{0.5\textwidth}
    \begin{align*}
      \text{reify}_t & : \text{NbE}_t(t) \to \text{expr}(t) \\
      \text{reify}_{t_1 \to t_2}(f) & \defeq \lambda v.\,\text{reify}_{t_2}(f(\text{reflect}_{t_1}(v))) \\
      \text{reify}_{b}(f) & \defeq f \\ \noalign{\vskip7pt}
      \text{reflect}_t & : \text{expr}(t) \to \text{NbE}_t(t) \\
      \text{reflect}_{t_1\to t_2}(e) & \defeq \lambda x.\,\text{reflect}_{t_2}(e(\text{reify}_{t_1}(x)) \\
      \text{reflect}_{b}(e) & \defeq e \\ \noalign{\vskip7pt}
    \end{align*}
  \end{subfigure}
  \begin{subfigure}[b]{0.5\textwidth}
    \begin{align*}
      \text{reduce} & : \text{expr}(t) \to \text{NbE}_t(t) \\
      \text{reduce}(\lambda v. \; e) & \defeq \lambda x. \; \text{reduce}([x/v]e) \\
      \text{reduce}(e_1~e_2) & \defeq \left(\text{reduce}(e_1)\right)(\text{reduce}(e_2)) \\
      \text{reduce}(x) & \defeq x \\
      \text{reduce}(c) & \defeq \text{reflect}(c) \\ \noalign{\vskip7pt}
      \text{NbE} & : \text{expr}(t) \to \text{expr}(t) \\
      \text{NbE}(e) & \defeq \text{reify}(\text{reduce}(e))
    \end{align*}
  \end{subfigure}
\caption{\label{fig:nbe}Implementation of normalization by evaluation}
\end{figure}

Now the core of NbE, shown in \autoref{fig:nbe}, is a pair of dual functions reify and reflect, for converting back and forth between syntax and semantics of the object language, defined by primitive recursion on type syntax.
We split out analysis of term syntax in a separate function reduce, defined by primitive recursion on term syntax, when usually this functionality would be mixed in with reflect.
The reason for this choice will become clear when we extend NbE.

We write $v$ for object-language variables and $x$ for metalanguage (Coq) variables, and we overload $\lambda$ notation using the metavariable kind to signal whether we are building a host $\lambda$ or a $\lambda$ syntax tree for the embedded language.
The crucial first clause for reduce replaces object-language variable $v$ with fresh metalanguage variable $x$, and then we are somehow tracking that all free variables in an argument to reduce must have been replaced with metalanguage variables by the time we reach them.
We reveal in \autoref{sec:PHOAS} the encoding decisions that make all the above legitimate, but first let us see how to integrate use of the rewriting operation from the previous section.
To fuse NbE with rewriting, we only modify the constant case of \texttt{reduce}.
First, we bind our specialized decision-tree engine (which rewrites \emph{at the root of an AST only}) under the name rewrite-head.

In the constant case, we still reflect the constant, but underneath the binders introduced by full $\eta$-expansion, we perform one instance of rewriting.
In other words, we change this one function-definition clause:
\begin{align*}
  \text{reflect}_{b}(e) & \defeq \text{rewrite-head}(e)
\end{align*}

It is important to note that a constant of function type will be $\eta$-expanded only once for each syntactic occurrence in the starting term, though the expanded function is effectively a thunk, waiting to perform rewriting again each time it is called.
From first principles, it is not clear why such a strategy terminates on all possible input terms.

The details so far are essentially the same as in the approach of \citet{Aehlig}.
Recall that their rewriter was implemented in a deeply embedded ML, while ours is implemented in Coq's logic, which enforces termination of all functions.
Aehlig et al.\ did not prove termination, which indeed does not hold for their rewriter in general, which works with untyped terms, not to mention the possibility of divergent rule-specific ML functions.
In contrast, we need to convince Coq up-front that our interleaved $\lambda$-term normalization and algebraic simplification always terminate.
Additionally, we must prove that rewriting preserves term denotations, which can easily devolve into tedious binder bookkeeping.

The next section introduces the techniques we use to avoid explicit termination proof or binder bookkeeping, in the context of a more general analysis of scaling challenges.

\section{Scaling Challenges}\label{sec:scaling}

\citet{Aehlig} only evaluated their implementation against closed programs.
What happens when we try to apply the approach to partial-evaluation problems that should generate thousands of lines of low-level code?

\subsection{Variable Environments Will Be Large}\label{sec:PHOAS}
We should think carefully about representation of ASTs, since many primitive operations on variables will run in the course of a single partial evaluation.
For instance, \citet{Aehlig} reported a significant performance improvement changing variable nodes from using strings to using de Bruijn indices~\cite{debruijn1972}.
However, de Bruijn indices and other first-order representations remain painful to work with.
We often need to fix up indices in a term being substituted in a new context.
Even looking up a variable in an environment tends to incur linear time overhead, thanks to traversal of a list.
Perhaps we can do better with some kind of balanced-tree data structure, but there is a fundamental performance gap versus the arrays that can be used in imperative implementations.
Unfortunately, it is difficult to integrate arrays soundly in a logic.
Also, even ignoring performance overheads, tedious binder bookkeeping complicates proofs.

Our strategy is to use a variable encoding that pushes all first-order bookkeeping off on Coq's kernel or the implementation of the language we extract to, which are themselves performance-tuned with some crucial pieces of imperative code.
Parametric higher-order abstract syntax (PHOAS)~\cite{PhoasICFP08} is a dependently typed encoding of syntax where binders are managed by the enclosing type system.
It allows for relatively easy implementation and proof for NbE, so we adopted it for our framework.

Here is the actual inductive definition of term syntax for our object language, PHOAS-style.
The characteristic oddity is that the core syntax type \texttt{expr} is parameterized on a dependent type family for representing variables.
However, the final representation type \texttt{Expr} uses first-class polymorphism over choices of variable type, bootstrapping on the metalanguage's parametricity to ensure that a syntax tree is agnostic to variable type.
\begin{minted}[fontsize=\small]{coq}
Inductive type := arrow (s d : type) | base (b : base_type).
Infix "→" := arrow.
Inductive expr (var : type -> Type) : type -> Type :=
| Var {t} (v : var t) : expr var t
| Abs {s d} (f : var s -> expr var d) : expr var (s → d)
| App {s d} (f : expr var (s → d)) (x : expr var s) : expr var d
| LetIn {a b} (x : expr var a) (f : var a -> expr var b) : expr var b
| Const {t} (c : const t) : expr var t.
Definition Expr (t : type) : Type := forall var, expr var t.
\end{minted}

A good example of encoding adequacy is assigning a simple denotational semantics.
First, a simple recursive function assigns meanings to types.
\begin{minted}[fontsize=\small]{coq}
Fixpoint denoteT (t : type) : Type := match t with
  | arrow s d => denoteT s -> denoteT d
  | base b    => denote_base_type b  end.
\end{minted}

Next we see the convenience of being able to \emph{use} an expression by choosing how it should represent variables.
Specifically, it is natural to choose \emph{the type-denotation function itself} as variable representation.
Especially note how this choice makes rigorous last section's convention (e.g., in the suspicious function-abstraction clause of reduce), where a recursive function enforces that values have always been substituted for variables early enough.
\begin{minted}[fontsize=\small]{coq}
Fixpoint denoteE {t} (e : expr denoteT t) : denoteT t := match e with
  | Var v     => v
  | Abs f     => λ x, denoteE (f x)
  | App f x   => (denoteE f) (denoteE x)
  | LetIn x f => let xv := denoteE x in denoteE f xv
  | Ident c   => denoteI c  end.
Definition DenoteE {t} (E : Expr t) : denoteT t := denoteE (E denoteT).
\end{minted}

It is now easy to follow the same script in making our rewriting-enabled NbE fully formal, in \autoref{fig:nbe2}.
Note especially the first clause of \texttt{reduce}, where we avoid variable substitution precisely because we have chosen to represent variables with normalized semantic values.
The subtlety there is that base-type semantic values are themselves expression syntax trees, which depend on a nested choice of variable representation, which we retain as a parameter throughout these recursive functions.
The final definition $\lambda$-quantifies over that choice.

\begin{figure}
  \begin{subfigure}[b]{0.5\textwidth}
    \begin{minted}[fontsize=\small]{coq}
Fixpoint nbeT var (t : type) : Type :=
match t with
| arrow s d => nbeT var s -> nbeT var d
| base b    => expr var b
end.


Fixpoint reify {var t}
  : nbeT var t -> expr var t :=
match t with
| arrow s d => λ f, Abs (λ x,
    reify (f (reflect (Var x))))
| base b    => λ e, e            end
    \end{minted}
  \end{subfigure}
  \begin{subfigure}[b]{0.5\textwidth}
    \begin{minted}[fontsize=\small]{coq}
with reflect{var t}:expr var t->nbeT var t
  := match t with
| arrow s d => λ e, λ x,
  reflect (App e (reify x))
| base b    => rewrite_head      end.
Fixpoint reduce{var t}(e:expr (nbeT var) t)
  : nbeT var t := match e with
| Abs e     => λ x, reduce (e (Var x))
| App e1 e2 => (reduce e1) (reduce e2)
| Var x     => x
| Ident c   => reflect (Ident c) end.
Definition Rewrite {t} (E:Expr t) : Expr t
  := λ var, reify (reduce (E (nbeT var t))).
    \end{minted}
  \end{subfigure}
  \caption{\label{fig:nbe2}PHOAS implementation of normalization by evaluation}
\end{figure}

One subtlety hidden in \autoref{fig:nbe2} in implicit arguments is in the final clause of \texttt{reduce}, where the two applications of the \texttt{Ident} constructor use different variable representations.
With all those details hashed out, we can prove a pleasingly simple correctness theorem, with a lemma for each main definition, with inductive structure mirroring recursive structure of the definition, also appealing to correctness of last section's pattern-compilation operations.
(We now use syntax $\llbracket \cdot \rrbracket$ for calls to \texttt{DenoteE}.)
$$\forall t, E : \textmintinline{coq}{Expr t}. \; \llbracket \textmintinline{coq}{Rewrite}(E) \rrbracket = \llbracket E \rrbracket$$


To understand how we now apply the soundness theorem in a tactic, it is important to note how the Coq kernel builds in reduction strategies.
These strategies have, to an extent, been tuned to work well to show equivalence between a simple denotational-semantics application and the semantic value it produces.
In contrast, it is rather difficult to code up one reduction strategy that works well for all partial-evaluation tasks.
Therefore, we should restrict ourselves to (1) running full reduction in the style of functional-language interpreters and (2) running normal reduction on ``known-good'' goals like correctness of evaluation of a denotational semantics on a concrete input.

Operationally, then, we apply our tactic in a goal containing a term $e$ that we want to partially evaluate.
In standard proof-by-reflection style, we \emph{reify} $e$ into some $E$ where $\llbracket E \rrbracket = e$, replacing $e$ accordingly, asking Coq's kernel to validate the equivalence via standard reduction.
Now we use the \mintinline{coq}{Rewrite} correctness theorem to replace $\llbracket E \rrbracket$ with $\llbracket \textmintinline{coq}{Rewrite}(E) \rrbracket$.
Next we ask the Coq kernel to simplify $\textmintinline{coq}{Rewrite}(E)$ by \emph{full reduction via native compilation}.
Finally, where $E'$ is the result of that reduction, we simplify $\llbracket E' \rrbracket$ with standard reduction.

We have been discussing representation of bound variables.
Also important is representation of constants (e.g., library functions mentioned in rewrite rules).
They could also be given some explicit first-order encoding, but dispatching on, say, strings or numbers for constants would be rather inefficient in our generated code.
Instead, we chose to have our Coq plugin generate a custom inductive type of constant codes, for each rewriter that we ask it to build with \texttt{Make}.
As a result, dispatching on a constant can happen in constant time, based on whatever pattern-matching is built into the execution language (either the Coq kernel or the target language of extraction).
To our knowledge, no past verified reduction tool in a proof assistant has employed that optimization.


\subsection{Subterm Sharing Is Crucial}\label{sec:under-lets}

For some large-scale partial-evaluation problems, it is important to represent output programs with sharing of common subterms.
Redundantly inlining shared subterms can lead to exponential increase in space requirements.
Consider the Fiat Cryptography~\cite{FiatCryptoSP19} example of generating a 64-bit implementation of field arithmetic for the P-256 elliptic curve.
The library has been converted manually to continuation-passing style, allowing proper generation of \mintinline{coq}{let} binders, whose variables are often mentioned multiple times.
We ran that old code generator (actually just a subset of its functionality, but optimized by us a bit further, as explained in \autoref{sec:macro}) on the P-256 example and found it took about 15 seconds to finish.
Then we modified reduction to inline \mintinline{coq}{let} binders instead of preserving them, at which point the job terminated with an out-of-memory error, on a machine with 64 GB of RAM.

We see a tension here between performance and niceness of library implementation.
When we built the original Fiat Cryptography, we found it necessary to CPS-convert the code to coax Coq into adequate reduction performance.
Then all of our correctness theorems were complicated by reasoning about continuations.
In fact, the CPS reasoning was so painful that at one point most algorithms in the template library were defined twice, once in continuation-passing style and once in direct-style code, because it was easier to prove the two equivalent and work with the non-CPS version than to reason about the CPS version directly.
It feels like a slippery slope on the path to implementing a domain-specific compiler, rather than taking advantage of the pleasing simplicity of partial evaluation on natural functional programs.
Our reduction engine takes shared-subterm preservation seriously while applying to libraries in direct style.

Our approach is \mintinline{coq}{let}-lifting: we lift \mintinline{coq}{let}s to top level, so that applications of functions to \mintinline{coq}{let}s are available for rewriting.
For example, we can perform the rewriting
\begin{align*}
  \texttt{map}\ (\lambda x.\, y+x)\ (\letin[{z:=e}{[0;1;z+1]}]) & 
  \; \leadsto \;
  \letin[{z:=e}{[y;y+1;y+(z+1)]}]
\end{align*}
using the rules
\begin{align*}
  \texttt{map}\ {?f}\ [] & \to []
  &
  \texttt{map}\ {?f}\ ({?x} :: {?xs}) & \to f\ x :: \texttt{map}\ f\ xs
  & {?n} + 0 & \to n 
\end{align*}

We define a telescope-style type family called \mintinline{coq}{UnderLets}:
\begin{minted}[fontsize=\small]{coq}
Inductive UnderLets {var} (T : Type) := Base (v : T)
| UnderLet {A} (e : @expr var A) (f : var A -> UnderLets T).
\end{minted}
A value of type \mintinline{coq}{UnderLets T} is a series of \texttt{let} binders (where each expression \mintinline{coq}{e} may mention earlier-bound variables) ending in a value of type \mintinline{coq}{T}.

Recall that the NbE type interpretation mapped base types to expression syntax trees.
We add flexibility, parameterizing by a Boolean declaring whether to introduce telescopes.

\begin{minted}[fontsize=\small]{coq}
Fixpoint nbeT' {var} (with_lets : bool) (t : type) := match t with
  | base t => if with_lets then @UnderLets var (@expr var t) else @expr var t
  | arrow s d => nbeT' false s -> nbeT' true d  end.
Definition nbeT := nbeT' false.      Definition nbeT_with_lets := nbeT' true.
\end{minted}


There are cases where naive preservation of \texttt{let} binders blocks later rewrites from triggering and leads to suboptimal performance, so we include some heuristics.
For instance, when the expression being bound is a constant, we always inline.
When the expression being bound is a series of list ``cons'' operations, we introduce a name for each individual list element, since such a list might be traversed multiple times in different ways.

\subsection{Rules Need Side Conditions}\label{sec:side-conditions}

Many useful algebraic simplifications require side conditions.
For example, bit-shifting operations are faster than divisions, so we might want a rule such as
\begin{align*}
  {?n} / {?m} & \to n \gg \log_2 m\text{\quad if\quad }2^{\lfloor \log_2 m \rfloor} = m
\end{align*}

The trouble is how to support predictable solving of side conditions during partial evaluation, where we may be rewriting in open terms.
We decided to sidestep this problem by allowing side conditions only as executable Boolean functions, to be applied only to variables that are confirmed as \emph{compile-time constants}, unlike \citet{rtac} who support general unification variables.
We added a variant of pattern variable that only matches constants.
Semantically, this variable style has no additional meaning, and in fact we implement it as a special identity function (notated as an apostrophe) that should be called in the right places within Coq lemma statements.
Rather, use of this identity function triggers the right behavior in our tactic code that reifies lemma statements.
\label{sec:explain-'}

Our reification inspects the hypotheses of lemma statements, using type classes to find decidable realizations of the predicates that are used, thereby synthesizing one Boolean expression of our deeply embedded term language, which stands for a decision procedure for the hypotheses.
The \mintinline{coq}{Make} command fails if any such expression contains pattern variables not marked as constants.

Hence, we encode the above rule as $\forall n, m. \; 2^{\lfloor \log_2(\texttt{'}m)\rfloor} = \texttt{'}m \to n / \texttt{'}m = n \gg \texttt{'}(\log_2 m)$.

\subsection{Side Conditions Need Abstract Interpretation}\label{sec:abs-int}

With our limitation that side conditions are decided by executable Boolean procedures, we cannot yet handle directly some of the rewrites needed for realistic compilation.
For instance, Fiat Cryptography reduces high-level functional to low-level code that only uses integer types available on the target hardware.
The starting library code works with arbitrary-precision integers, while the generated low-level code should be careful to avoid unintended integer overflow.
As a result, the setup may be too naive for our running example rule ${?n} + 0 \to n$.
When we get to reducing fixed-precision-integer terms, we must be legalistic:
\begin{align*}
  \texttt{add\_with\_carry}_{64}({?n}, 0) & \to (0, n)\text{\ \ if\ \ }0 \le n < 2^{64}
\end{align*}

We developed a design pattern to handle this kind of rule.

First, we introduce a family of functions $\texttt{clip}_{l,u}$, each of which forces its integer argument to respect lower bound $l$ and upper bound $u$.
Partial evaluation is proved with respect to unknown realizations of these functions, only requiring that $\texttt{clip}_{l, u}(n) = n$ when $l \leq n < u$.
Now, before we begin partial evaluation, we can run a verified abstract interpreter to find conservative bounds for each program variable.
When bounds $l$ and $u$ are found for variable $x$, it is sound to replace $x$ with $\texttt{clip}_{l,u}(x)$.
Therefore, at the end of this phase, we assume all variable occurrences have been rewritten in this manner to record their proved bounds.

Second, we proceed with our example rule refactored:
\begin{align*}
  \texttt{add\_with\_carry}_{64}(\texttt{clip}_{\texttt{'}{?l},\texttt{'}{?u}}({?n}), 0) & \to (0, \texttt{clip}_{l,u}(n)) 
  \text{\ \ if\ \ }u < 2^{64}
\end{align*}
If the abstract interpreter did its job, then all lower and upper bounds are constants, and we can execute side conditions straightforwardly during pattern matching.

See \autoref{sec:implementation-and-usage}\ifthenelse{\boolean{tobesplit}}{%
  \footnote{%
  The appendix of this paper is available online at \url{https://jasongross.github.io/papers/2022-rewriting-itp-supplement.pdf}.%
  }%
}{}
for discussion of some further twists in the implementation.

\section{Evaluation}\label{sec:evaluation}

Our implementation, available on GitHub at \githublink{mit-plv}{rewriter}[ITP-2022-perf-data] and with a roadmap in \autoref{sec:CodeSupplement-more}, includes a mix of Coq code for the proved core of rewriting, tactic code for setting up proper use of that core, and OCaml plugin code for the manipulations beyond the tactic language's current capabilities.
We report here on evidence that the tool is effective, first in terms of productivity by users and then in terms of compile-time performance.

\subsection{Iteration on the Fiat Cryptography Compiler}\label{sec:iteration}

We ported Fiat Cryptography's core compiler functionality to use our framework.
The result is now used in production by a number of open-source projects.
We were glad to retire the CPS versions of verified arithmetic functions, which had been present only to support predictable reduction with subterm sharing.
More importantly, it became easy to experiment with new transformations via proving new rewrite theorems, directly in normal Coq syntax, including the following, all justified by demand from real users:
\begin{itemize}
\item Reassociating arithmetic to minimize the bitwidths of intermediate results
\item Multiplication primitives that separately return high halves and low halves
\item Strings and a ``comment'' function of type $\forall A. \; \texttt{string} \to A \to A$
\item Support for bitwise exclusive-or
\item A special marker to block C compilers from introducing conditional jumps in code that should be constant-time
\item Eliding bitmask-with-constant operations that can be proved as no-ops
\item Rules to introduce conditional moves (on supported platforms)
\item New hardware backend, via rules that invoke special instructions of a cryptographic accelerator
\item New hardware backend, with a requirement that all intermediate integers have the same bitwidth, via rules to break wider operations down into several narrower operations
\end{itemize}




\subsection{Microbenchmarks}\label{sec:micro}

\def\NoBindersSubfloatNval{3}%
\def\NoBindersSubfloatXRow{\thisrow{param-2-m}*(2^(\thisrow{param-1-n}+1)-1)}%

\begin{figure*}
  \newsavebox{\NestedBindersSubfloat}%
  \sbox{\NestedBindersSubfloat}{%
    \adjustbox{valign=t}{\resizebox{0.32\textwidth}{!}{\beginTikzpictureStamped[only marks]{
      \input{perf-UnderLetsPlus0.csv.md5} \space
    }
      \pgfplotsset{every axis legend/.append style={
          at={(0.5,-0.2)},
          anchor=north}}
      \begin{axis}[xlabel=\# of let binders,
          ylabel=time (s),
          scaled x ticks=false,
          ymax=65,
          xmax=5000,
          table/col sep=comma,
          table/x=param-n
        ]
        \addplot[mark=o,color=red]         table[y=rewrite-strat(bottomup)-regression-exponential-user]{perf-UnderLetsPlus0.csv};
        \addplot[mark=triangle,color=red]  table[y=rewrite-strat(topdown)-regression-exponential-user] {perf-UnderLetsPlus0.csv};
        \addplot[mark=square,color=red]    table[y=setoid-rewrite-regression-cubic-user]               {perf-UnderLetsPlus0.csv};
        \addplot[mark=+,color=blue]        table[y=Rewrite-for-gen-user]                               {perf-UnderLetsPlus0.csv};
        \addplot[mark=x,color=ForestGreen] table[y=rewriting-user]                                     {perf-UnderLetsPlus0.csv};
        \legend{rewrite\_strat bottomup,rewrite\_strat topdown,setoid\_rewrite,{Our approach including reification, cbv, etc.},Our approach (only rewriting)}
      \end{axis}
    \end{tikzpicture}}}}%
  \newsavebox{\BindersAndRecursiveFunctionsSubfloat}%
  \sbox{\BindersAndRecursiveFunctionsSubfloat}{%
    \adjustbox{valign=t}{\resizebox{0.32\textwidth}{!}{\beginTikzpictureStamped[only marks]{
        \input{perf-UnderLetsPlus0.csv.md5} \space
    }
      \pgfplotsset{every axis legend/.append style={
          at={(0.5,-0.2)},
          anchor=north}}
      \begin{axis}[xlabel=$n\cdot m$,
          ylabel=time (s),
          scaled x ticks=false,
          ymax=27,
          xmax=12000,
          table/col sep=comma,
          table/x=param-0-nm]
        \addplot[mark=o,color=red]         table[y=rewrite-strat(bottomup-bottomup)-regression-exponential-user]{perf-LiftLetsMap.csv};
        \addplot[mark=triangle,color=red]  table[y=rewrite-strat(topdown-bottomup)-regression-exponential-user]{perf-LiftLetsMap.csv};
        \addplot[mark=square,color=red]    table[y=setoid-rewrite-regression-cubic-user]{perf-LiftLetsMap.csv};
        \addplot[mark=+,color=blue]        table[y=Rewrite-for-gen-user]{perf-LiftLetsMap.csv};
        \addplot[mark=x,color=ForestGreen] table[y=rewriting-user]{perf-LiftLetsMap.csv};
        \addplot[mark=*,color=purple]      table[y=cps+vm-compute-regression-quadratic-user]{perf-LiftLetsMap.csv};
        \legend{rewrite\_strat bottomup,rewrite\_strat topdown,repeat setoid\_rewrite,{Our approach including reification, cbv, etc.},Our approach (only rewriting),cps+vm\_compute}
      \end{axis}
    \end{tikzpicture}}}}%
  \newsavebox{\NoBindersSubfloat}%
  \sbox{\NoBindersSubfloat}{%
    \edef\nval{\NoBindersSubfloatNval}%
    \adjustbox{valign=t}{\resizebox{0.32\textwidth}{!}{\beginTikzpictureStamped[only marks]{
        \input{perf-Plus0Tree.csv.md5} \space
        \nval
    }
      \pgfplotsset{every axis legend/.append style={
          at={(0.5,-0.2)},
          anchor=north}}
      \begin{axis}[xlabel=\# of rewrite locations,
          scaled x ticks=false,
          ylabel=time (s),
          ymax=7,
          xmax=15000,
          xtick distance=3000,
          table/col sep=comma,
          table/x expr={\NoBindersSubfloatXRow}]
        \addplot[discard if not={param-1-n}{\nval},mark=square,color=red] table[y=rewrite-strat(bottomup)-regression-cubic-user]{perf-Plus0Tree.csv};
        \addplot[discard if not={param-1-n}{\nval},mark=*,color=red] table[y=setoid-rewrite-regression-cubic-user]{perf-Plus0Tree.csv};
        \addplot[discard if not={param-1-n}{\nval},mark=triangle,color=red] table[y=rewrite-strat(topdown)-regression-cubic-user]{perf-Plus0Tree.csv};
        \addplot[discard if not={param-1-n}{\nval},mark=o,color=red] table[y=rewrite!-regression-cubic-user]{perf-Plus0Tree.csv};
        \addplot[discard if not={param-1-n}{\nval},mark=+,color=blue] table[y=Rewrite-for-gen-user]{perf-Plus0Tree.csv};
        \addplot[discard if not={param-1-n}{\nval},mark=x,color=ForestGreen] table[y=rewriting-user]{perf-Plus0Tree.csv};
        \legend{rewrite\_strat bottomup,setoid\_rewrite,rewrite\_strat topdown,rewrite!,{Our approach including reification, cbv, etc.},Our approach (only rewriting)}
      \end{axis}
    \end{tikzpicture}}}}%
  \newsavebox{\FiatCryptoSubfloat}%
  \sbox{\FiatCryptoSubfloat}{%
    \adjustbox{valign=t}{\resizebox{0.32\textwidth}{!}{\beginTikzpictureStamped[only marks]{
      \input{perf-fiat-crypto.csv.md5} \space
    }
      \pgfplotsset{every axis legend/.append style={
          at={(0.5,-0.2)},
          anchor=north}}
      \begin{axis}[xlabel=prime,ylabel=time relative to original Fiat Crypto,xmode=log, ymode=log,log basis x={2},
        ]
        \addplot[mark=*,color=red] table[x=prime,y=OldSynthesisAndPackage-over-OldSynthesisAndPackage-real,col sep=comma]{perf-fiat-crypto.csv};
        \addlegendentry{Original Fiat Crypto (includes reification+rewriting)}
        \addplot[mark=+,color=blue] table[x=prime,y=NewVMFull-over-OldSynthesisAndPackage-real,col sep=comma]{perf-fiat-crypto.csv};
        \addlegendentry{Our approach w/ Coq's VM}


        \addplot[mark=x,color=ForestGreen] table[x=prime,y=NewExtractionFull-over-OldSynthesisAndPackage-real,col sep=comma]{perf-fiat-crypto.csv};
        \addlegendentry{Our approach w/ extracted OCaml}
      \end{axis}
    \end{tikzpicture}}}}%
  \newcommand{\vphantomSubfloatOne}{\vphantom{\usebox{\NestedBindersSubfloat}}\vphantom{\usebox{\NoBindersSubfloat}}}%
  \newcommand{\vphantomSubfloatTwo}{\vphantom{\usebox{\BindersAndRecursiveFunctionsSubfloat}}\vphantom{\usebox{\FiatCryptoSubfloat}}}%
  \subfloat[No binders]{\usebox{\NoBindersSubfloat}\vphantomSubfloatOne\label{fig:timing-Plus0Tree}}%
  \quad
  \subfloat[Nested binders]{\usebox{\NestedBindersSubfloat}\vphantomSubfloatOne\label{fig:timing-UnderLetsPlus0}}%
  \quad
  \subfloat[Fiat Cryptography]{\usebox{\FiatCryptoSubfloat}\vphantomSubfloatTwo\label{fig:scaling}}

  \caption{Timing of different partial-evaluation implementations}\label{fig:multi-timing}
\end{figure*}


Now we turn to evaluating performance of generated compilers.
We start with microbenchmarks focusing attention on particular aspects of reduction and rewriting, with \autoref{sec:additionalMicro} going into more detail, including on a few more benchmarks.

Our first example family, \emph{nested binders}, has two integer parameters $n$ and $m$.
An expression tree is built with $2^n$ copies of an expression, which is itself a free variable with $m$ ``useless'' additions of zero.
We want to see all copies of this expression reduced to just the variable.
\Vref{fig:timing-Plus0Tree} shows the results for $n = \NoBindersSubfloatNval$ as we scale $m$.
The comparison points are Coq's \texttt{rewrite!}, \texttt{setoid\_rewrite}, and \texttt{rewrite\_strat}.
The first two perform one rewrite at a time, taking minimal advantage of commonalities across them and thus generating quite large, redundant proof terms.
The third makes top-down or bottom-up passes with combined generation of proof terms.
For our own approach, we list both the total time and the time taken for core execution of a verified rewrite engine, without counting reification (converting goals to ASTs) or its inverse (interpreting results back to normal-looking goals).
The comparison here is very favorable for our approach so long as $m > 2$.
(See \autoref{sec:additionalPlots:Plus0Tree} 
for more detailed plots.)

Now consider what happens when we use \mintinline{coq}{let} binders to share subterms within repeated addition of zero, incorporating exponentially many additions with linearly sized terms.
\Vref{fig:timing-UnderLetsPlus0} shows the results.
The comparison here is again very favorable for our approach.
The competing tactics spike upward toward timeouts at just a few hundred generated binders, while our engine is only taking about 10 seconds for examples with 5,000 nested binders.

Although we have made our comparison against the built-in tactics \mintinline{coq}{setoid_rewrite} and \mintinline{coq}{rewrite_strat}, by analyzing the performance in detail, we can argue that these performance bottlenecks are likely to hold for any proof assistant designed like Coq.
Detailed debugging reveals five performance bottlenecks in the existing tactics, discussed in \autoref{sec:setoid-rewrite-bottlenecks}.

\subsection{Macrobenchmark: Fiat Cryptography}\label{sec:macro}

Finally, we consider an experiment (described in more detail in \autoref{sec:additionalMacro}) replicating the generation of performance-competitive finite-field-arithmetic code for all popular elliptic curves by \citet{FiatCryptoSP19}.
In all cases, we generate essentially the same code as they did, so we only measure performance of the code-generation process.
We stage partial evaluation with three different reduction engines (i.e., three \mintinline{coq}{Make} invocations), respectively applying 85, 56, and 44 rewrite rules (with only 2 rules shared across engines), taking total time of about 5 minutes to generate all three engines.
These engines support 95 distinct function symbols.

\Vref{fig:scaling} graphs running time of three different partial-evaluation and rewriting methods for Fiat Cryptography, as the prime modulus of arithmetic scales up.
Times are normalized to the performance of the original method of \citet{FiatCryptoSP19}, which relied on standard Coq reduction to evaluate code that had been manually written in CPS, followed by reification and a custom ad-hoc simplification and rewriting engine.

As the figure shows, our approach gives about a 10$\times$--1000$\times$ speed-up over the original Fiat Cryptography pipeline.
Inspection of the timing profiles of the original pipeline reveals that reification dominates the timing profile; since partial evaluation is performed by Coq's kernel, reification must happen \emph{after} partial evaluation, and hence the size of the term being reified grows with the size of the output code.
Also recall that the old approach required rewriting Fiat Cryptography's library of arithmetic functions in continuation-passing style, enduring this complexity in library correctness proofs, while our new approach applies to a direct-style library.
Finally, the old approach included a custom reflection-based arithmetic simplifier for term syntax, run after traditional reduction, whereas now we are able to apply a generic engine that combines both, without requiring more than proving traditional rewrites.

The figure also confirms a clear performance advantage of running reduction in code extracted to OCaml, which is possible because our plugin produces verified code in Coq's functional language.
The extracted version is about 10$\times$ faster than running in Coq's kernel.

\section{Future Work}

By far the biggest next step for our engine is to integrate abstract interpretation with rewriting and partial evaluation.
We expect this would net us asymptotic performance gains
as described in \autoref{sec:fusing-compiler-passes}.
Additionally, it would allow us to simplify the phrasing of many of our post-abstract-interpretation rewrite rules, by relegating bounds information to side conditions rather than requiring that they appear in the syntactic form of the rule.

There are also a number of natural extensions to our engine.
For instance, we do not yet allow pattern variables marked as ``constants only'' to apply to container datatypes; we limit the mixing of higher-order and polymorphic types, as well as limiting use of first-class polymorphism; we do not support rewriting with equalities of nonfully-applied functions; we only support decidable predicates as rule side conditions, and the predicates may only mention pattern variables restricted to matching constants; we have hardcoded support for a small set of container types and their eliminators; we support rewriting with equality and no other relations
; and we require decidable equality for all types mentioned in rules.

\clearpage
\bibliography{rewriting}
\clearpage

\startsupplement
\appendix

\FloatBarrier\section{Performance Bottlenecks of Proof-Producing Rewriting}\label{sec:setoid-rewrite-bottlenecks}

Although we have made our performance comparison against the built-in Coq tactics \mintinline{coq}{setoid_rewrite} and \mintinline{coq}{rewrite_strat}, by analyzing the performance in detail, we can argue that these performance bottlenecks are likely to hold for any proof assistant designed like Coq.
Detailed debugging reveals five performance bottlenecks in the existing rewriting tactics.

\subsection{Bad performance scaling in sizes of existential-variable contexts}

We found that even when there are no occurrences fully matching the rule, \mintinline{coq}{setoid_rewrite} can still be \emph{cubic} in the number of binders (or, more accurately, quadratic in the number of binders with an additional multiplicative linear factor of the number of head-symbol matches).
Rewriting without any successful matches takes nearly as much time as \mintinline{coq}{setoid_rewrite} in this microbenchmark; by the time we are looking at goals with 400 binders, the difference is less than 5\%.

We posit that this overhead comes from \mintinline{coq}{setoid_rewrite} looking for head-symbol matches and then creating evars (existential variables) to instantiate the arguments of the lemmas for each head-symbol-match location; hence even if there are no matches of the rule as a whole, there may still be head-symbol matches.
Since Coq uses a locally nameless representation~\cite{LocallyNameless} for its terms, evar contexts are necessarily represented as \emph{named} contexts.
Representing a substitution between named contexts takes linear space, even when the substitution is trivial, and hence each evar incurs overhead linear in the number of binders above it.
Furthermore, fresh-name generation in Coq is quadratic in the size of the context, and since evar-context creation uses fresh-name generation, the additional multiplicative factor likely comes from fresh-name generation.
(Note, though, that this pattern suggests that the true performance is quartic rather than merely cubic.
However, doing a linear regression on a $\log$-$\log$ of the data suggests that the performance is genuinely cubic rather than quartic.)

Note that this overhead is inherent to the use of a locally nameless term representation.
To fix it, Coq would likely have to represent identity evar contexts using a compact representation, which is only naturally available for de Bruijn representations.
Any rewriting system that uses unification variables with a locally nameless (or named) context will incur at least quadratic overhead on this benchmark.

Note that \mintinline{coq}{rewrite_strat} uses exactly the same rewriting engine as \mintinline{coq}{setoid_rewrite}, just with a different strategy.
We found that \mintinline{coq}{setoid_rewrite} and \mintinline{coq}{rewrite_strat} have identical performance when there are no matches and generate identical proof terms when there are matches.
Hence we can conclude that the difference in performance between \mintinline{coq}{rewrite_strat} and \mintinline{coq}{setoid_rewrite} is entirely due to an increased number of failed rewrite attempts.

\subsection{Proof-term size}

Setting aside the performance bottleneck in constructing the matches in the first place, we can ask the question: how much cost is associated to the proof terms?
One way to ask this question in Coq is to see how long it takes to run \mintinline{coq}{Qed}.
While \mintinline{coq}{Qed} time is asymptotically better, it is still quadratic in the number of binders.
This outcome is unsurprising, because the proof-term size is quadratic in the number of binders.
On this microbenchmark, we found that \mintinline{coq}{Qed} time hits one second at about 250 binders, and using the best-fit quadratic line suggests that it would hit 10 seconds at about 800 binders and 100 seconds at about 2\,500 binders.
While this may be reasonable for the microbenchmarks, which only contain as many rewrite occurrences as there are binders, it would become unwieldy to try to build and typecheck such a proof with a rule for every primitive reduction step, which would be required if we want to avoid manually CPS-converting the code in Fiat Cryptography.

The quadratic factor in the proof term comes because we repeat subterms of the goal linearly in the number of rewrites.
For example, if we want to rewrite \mintinline{coq}{f (f x)} into \mintinline{coq}{g (g x)} by the equation \mintinline{coq}{∀ x, f x = g x}, then we will first rewrite \mintinline{coq}{f x} into \mintinline{coq}{g x}, and then rewrite \mintinline{coq}{f (g x)} into \mintinline{coq}{g (g x)}.
Note that \mintinline{coq}{g x} occurs three times (and will continue to occur in every subsequent step).

\subsection{Poor subterm sharing}

How easy is it to share subterms and create a linearly sized proof?
While it is relatively straightforward to share subterms using \mintinline{coq}{let} binders when the rewrite locations are not under any binders, it is not at all obvious how to share subterms when the terms occur under different binders.
Hence any rewriting algorithm that does not find a way to share subterms across different contexts will incur a quadratic factor in proof-building and proof-checking time, and we expect this factor will be significant enough to make applications to projects as large as Fiat Crypto infeasible.

\subsection{Overhead from the \mintinline{coq}{let} typing rule}

Suppose we had a proof-producing rewriting algorithm that shared subterms even under binders.
Would it be enough?
It turns out that even when the proof size is linear in the number of binders, the cost to typecheck it in Coq is still quadratic!
The reason is that when checking that \texttt{f : T} in a context \mintinline{coq}{x := v}, to check that \mintinline{coq}{let x := v in f} has type \texttt{T} (assuming that \mintinline{coq}{x} does not occur in \texttt{T}), Coq will substitute \mintinline{coq}{v} for \mintinline{coq}{x} in \texttt{T}.
So if a proof term has $n$ \mintinline{coq}{let} binders (e.g., used for sharing subterms), Coq will perform $n$ substitutions on the type of the proof term, even if none of the \mintinline{coq}{let} binders are used.
If the number of \mintinline{coq}{let} binders is linear in the size of the type, there is quadratic overhead in proof-checking time, even when the proof-term size is linear.

We performed a microbenchmark on a rewriting goal with no binders (because there is an obvious algorithm for sharing subterms in that case) and found that the proof-checking time reached about one second at about 2\,000 binders and reached 10 seconds at about 7\,000 binders.
While these results might seem good enough for Fiat Cryptography, we expect that there are hundreds of thousands of primitive reduction/rewriting steps even when there are only a few hundred binders in the output term, and we would need \mintinline{coq}{let} binders for each of them.
Furthermore, we expect that getting such an algorithm correct would be quite tricky.

Fixing this quadratic bottleneck would, as far as we can tell, require deep changes in how Coq is implemented; it would either require reworking all of Coq to operate on some efficient representation of delayed substitutions paired with unsubstituted terms, or else it would require changing the typing rules of the type theory itself to remove this substitution from the typing rule for \mintinline{coq}{let}.
Note that there is a similar issue that crops up for function application and abstraction.

\subsection{Inherent advantages of reflection}

Finally, even if this quadratic bottleneck were fixed, \citet{Aehlig} reported a $10\times$--$100\times$ speed-up over the \emph{simp} tactic in Isabelle, which performs all of the intermediate rewriting steps via the kernel API.
Their results suggest that even if all of the superlinear bottlenecks were fixed---no small undertaking---rewriting and partial evaluation via reflection might still be orders of magnitude faster than any proof-term-generating tactic.

\clearpage

\section{Additional Benchmarking Plots}\label{sec:additionalPlots}

\subsection{Rewriting Without Binders}\label{sec:additionalPlots:Plus0Tree}

\begin{figure*}
\newcommand{\PlusZeroTreeMValSubfloat}[1]{%
  \def\mval{#1}%
  \subfloat[No binders ($m=\mval$)]{%
  \resizebox{0.4\textwidth}{!}{\beginTikzpictureStamped[only marks]{
    \input{perf-Plus0Tree.csv.md5} \space
    \mval
  }
    \pgfplotsset{every axis legend/.append style={
        at={(0.5,-0.2)},
        anchor=north}}
    \begin{axis}[xlabel=$n$,
        ylabel=time (s),
        ymode=log,
        ymin=0.002,
        ymax=110,
        xmax=17,
        table/col sep=comma,
        table/x=param-1-n]
      \addplot[discard if not={param-2-m}{\mval},mark=square,color=red] table[y=rewrite-strat(bottomup)-regression-cubic-user]{perf-Plus0Tree.csv};
      \addplot[discard if not={param-2-m}{\mval},mark=*,color=red] table[y=setoid-rewrite-regression-cubic-user]{perf-Plus0Tree.csv};
      \addplot[discard if not={param-2-m}{\mval},mark=triangle,color=red] table[y=rewrite-strat(topdown)-regression-cubic-user]{perf-Plus0Tree.csv};
      \addplot[discard if not={param-2-m}{\mval},mark=o,color=red] table[y=rewrite!-regression-cubic-user]{perf-Plus0Tree.csv};
      \addplot[discard if not={param-2-m}{\mval},mark=+,color=blue] table[y=Rewrite-for-gen-user]{perf-Plus0Tree.csv};
      \addplot[discard if not={param-2-m}{\mval},mark=x,color=ForestGreen] table[y=rewriting-user]{perf-Plus0Tree.csv};
      \legend{rewrite\_strat bottomup,setoid\_rewrite,rewrite\_strat topdown,rewrite!,{Our approach including reification, cbv, etc.},Our approach (only rewriting)}
    \end{axis}
  \end{tikzpicture}}%
 \label{fig:timing-Plus0Tree-m=\mval}}%
}%
\PlusZeroTreeMValSubfloat{1}\qquad
\PlusZeroTreeMValSubfloat{2}\qquad
\PlusZeroTreeMValSubfloat{3}
  \caption{Timing of different partial-evaluation implementations for code with no binders for fixed $m$.  Note that we have a logarithmic time scale, because term size is proportional to $2^n$.}\label{fig:timing-Plus0Tree-fixed-m}
\end{figure*}

\begin{figure*}
\newcommand{\PlusZeroTreeNValSubfloat}[1]{%
  \def\nval{#1}%
  \subfloat[No binders ($n=\nval$)]{%
  \resizebox{0.4\textwidth}{!}{\beginTikzpictureStamped[only marks]{
      \input{perf-Plus0Tree.csv.md5} \space
      \nval
  }
    \pgfplotsset{every axis legend/.append style={
        at={(0.5,-0.2)},
        anchor=north}}
    \begin{axis}[xlabel=$m$,
        ylabel=time (s),
        ymax=10,
        xmax=1100,
        table/col sep=comma,
        table/x=param-2-m]
      \addplot[discard if not={param-1-n}{\nval},mark=square,color=red] table[y=rewrite-strat(bottomup)-regression-cubic-user]{perf-Plus0Tree.csv};
      \addplot[discard if not={param-1-n}{\nval},mark=*,color=red] table[y=setoid-rewrite-regression-cubic-user]{perf-Plus0Tree.csv};
      \addplot[discard if not={param-1-n}{\nval},mark=triangle,color=red] table[y=rewrite-strat(topdown)-regression-cubic-user]{perf-Plus0Tree.csv};
      \addplot[discard if not={param-1-n}{\nval},mark=o,color=red] table[y=rewrite!-regression-cubic-user]{perf-Plus0Tree.csv};
      \addplot[discard if not={param-1-n}{\nval},mark=+,color=blue] table[y=Rewrite-for-gen-user]{perf-Plus0Tree.csv};
      \addplot[discard if not={param-1-n}{\nval},mark=x,color=ForestGreen] table[y=rewriting-user]{perf-Plus0Tree.csv};
      \legend{rewrite\_strat bottomup,setoid\_rewrite,rewrite\_strat topdown,rewrite!,{Our approach including reification, cbv, etc.},Our approach (only rewriting)}
    \end{axis}
  \end{tikzpicture}}%
 \label{fig:timing-Plus0Tree-n=\nval}}%
}%
\PlusZeroTreeNValSubfloat{1}\qquad
\PlusZeroTreeNValSubfloat{2}\qquad
\PlusZeroTreeNValSubfloat{3}\qquad
\PlusZeroTreeNValSubfloat{9}
  \caption{Timing of different partial-evaluation implementations for code with no binders for fixed $n$ (1, 2, 3, and then we jump to 9)}\label{fig:timing-Plus0Tree-fixed-n}
\end{figure*}


The code in \autoref{fig:micro:Plus0Tree:code} in \autoref{sec:micro:Plus0Tree} is parameterized on both $n$, the height of the tree, and $m$, the number of rewriting occurrences per node.
The plot in \autoref{fig:timing-Plus0Tree} displays only the case of $n=\NoBindersSubfloatNval$.
The plots in \autoref{fig:timing-Plus0Tree-fixed-m} display how performance scales as a factor of $n$ for fixed $m$, and the plots in \autoref{fig:timing-Plus0Tree-fixed-n} display how performance scales as a factor of $m$ for fixed $n$.
Note the logarithmic scaling on the time axis in the plots in \autoref{fig:timing-Plus0Tree-fixed-m}, as term size is proportional to $m\cdot 2^n$.

We can see from these graphs and the ones in \autoref{fig:timing-Plus0Tree-fixed-n} that
(a) we incur constant overhead over most of the other methods, which dominates on small examples;
(b) when the term is quite large and there are few opportunities for rewriting relative to the term size (i.e., $m \le 2$), we are worse than \mintinline{coq}{rewrite !Z.add_0_r} but still better than the other methods; and
(c) when there are many opportunities for rewriting relative to the term size ($m > 2$), we thoroughly dominate the other methods.

\clearpage

\subsection{Additional Information on the Fiat Cryptography Benchmark}\label{sec:additionalPlots:FiatCrypto}\label{sec:additionalMacro}

The data for this benchmark can be found on GitHub at \githublink{mit-plv}{fiat-crypto}[perf-testing-data-ITP-2022-rewriting].

\begin{figure*}[b]
  \newcommand{\MacroSubfloat}[3]{%
    \subfloat[Timing of different partial-evaluation implementations for Fiat Cryptography as prime modulus grows (only #2 #3)]{%
  \resizebox{0.45\textwidth}{!}{\beginTikzpictureStamped[only marks]{
      \input{perf-fiat-crypto.csv.md5} \space
  }
    \pgfplotsset{every axis legend/.append style={
        at={(0.5,-0.2)},
        anchor=north}}
    \begin{axis}[xlabel=prime,ylabel=time (s),xmode=log, ymode=log,log basis x={2}]
      \addplot[mark=*,color=red] table[x=prime,y=#1-#3-OldSynthesisAndPackage-real,col sep=comma]{perf-fiat-crypto.csv};
      \addlegendentry{Old approach (including reification+rewriting)}
      \addplot[mark=+,color=blue] table[x=prime,y=#1-#3-NewVMFull-real,col sep=comma]{perf-fiat-crypto.csv};
      \addlegendentry{Our approach w/ Coq's VM}


      \addplot[mark=x,color=ForestGreen] table[x=prime,y=#1-#3-NewExtractionFull-real,col sep=comma]{perf-fiat-crypto.csv};
      \addlegendentry{Our approach w/ extracted OCaml}
    \end{axis}
  \end{tikzpicture}}%
 \label{fig:timing--only-#1-#3}}%
  }%
  \MacroSubfloat{UnsaturatedSolinas}{unsaturated Solinas}{x32}\qquad
  \MacroSubfloat{UnsaturatedSolinas}{unsaturated Solinas}{x64}\\
  \MacroSubfloat{WordByWordMontgomery}{word-by-word Montgomery}{x32}\qquad
  \MacroSubfloat{WordByWordMontgomery}{word-by-word Montgomery}{x64}%
  \caption{\label{fig:timing-macro-various}Timing of different partial-evaluation implementations for Fiat Cryptography vs.\ prime modulus}
\end{figure*}

It may also be useful to see performance results with absolute times, rather than normalized execution ratios vs.\ the original Fiat Cryptography implementation.
Furthermore, the benchmarks fit into four quite different groupings: elements of the cross product of two algorithms (unsaturated Solinas and word-by-word Montgomery) and bitwidths of target architectures (32-bit or 64-bit).
Here we provide absolute-time graphs by grouping in \autoref{fig:timing-macro-various}.

\clearpage
\begin{figure}[t]
  \subfloat[Expressions computing initial code for Rewriting Without Binders]{%
  {\small $\begin{aligned}
  \text{iter}_m(v) & = v + \underbrace{0 + 0 + \cdots + 0}_m \\
  \text{tree}_{0,m}(v) &= \text{iter}_m(v + v) \\
  \text{tree}_{n+1,m}(v) &= \text{iter}_m(\text{tree}_{n,m}(v) + \text{tree}_{n,m}(v))
    \end{aligned}$}%
  \label{fig:micro:Plus0Tree:code}}
  \qquad
  \subfloat[Initial code for Rewriting Under Binders]{%
  {\small $\begin{aligned}
  & \letin[{v_1 := v_0 + v_0 + 0}{}] \\ \noalign{\vskip-7pt}
  & \vdots \\ \noalign{\vskip-3pt}
  & \letin[{v_n := v_{n-1} + v_{n-1} + 0}{}] \\
  & v_n + v_n + 0
\end{aligned}$}%
  \label{fig:micro:UnderLetsPlus0:code}%
  }
  \caption{\label{fig:micro:Plus0Tree+UnderLetsPlus0:code}Code for rewriting without and under binders}
\end{figure}

\FloatBarrier\section{Additional Information on Microbenchmarks}\label{sec:additionalMicro}

We performed all benchmarks on a 3.5 GHz Intel Haswell running Linux and Coq 8.11.1.
We name the subsections here with the names that show up in the code supplement.

\subsection{Rewriting Without Binders: Plus0Tree}\label{sec:Plus0Tree-more}\label{sec:micro:Plus0Tree}

Consider the code defined by the expression $\text{tree}_{n,m}(v)$ in \autoref{fig:micro:Plus0Tree:code}.
We want to remove all of the${}+0$s.
There are $\Theta(m \cdot 2^n)$ such rewriting locations.
We can start from this expression directly, in which case reification alone takes as much time as Coq's \texttt{rewrite}.
As the reification method was not especially optimized, and there exist fast reification methods~\cite{ReificationITP18}, we instead start from a call to a recursive function that generates such an expression.


We use two definitions for this microbenchmark:
\begin{minted}[fontsize=\small]{coq}
Definition iter (m : nat) (acc v : Z) :=
  @nat_rect (fun _ => Z -> Z)
    (fun acc => acc)
    (fun _ rec acc => rec (acc + v))
    m
    acc.
Definition make_tree (n m : nat) (v acc : Z) :=
 Eval cbv [iter] in
  @nat_rect (fun _ => Z * Z -> Z)
    (fun '(v, acc) => iter m (acc + acc) v)
    (fun _ rec '(v, acc) =>
      iter m (rec (v, acc) + rec (v, acc)) v)
    n
    (v, acc).
\end{minted}

\subsection{Rewriting Under Binders: UnderLetsPlus0}\label{sec:UnderLetsPlus0-more}\label{sec:micro:UnderLetsPlus0}

Consider now the code in \autoref{fig:micro:UnderLetsPlus0:code}, which is a version of the code above where redundant expressions are shared via \mintinline{coq}{let} bindings.

The code used to define this microbenchmark is
\begin{minted}[fontsize=\small]{coq}
Definition make_lets_def (n:nat) (v acc : Z) :=
 @nat_rect (fun _ => Z * Z -> Z)
   (fun '(v, acc) => acc + acc + v)
   (fun _ rec '(v, acc) =>
     dlet acc := acc + acc + v in rec (v, acc))
   n
   (v, acc).
\end{minted}
We note some details of the rewriting framework that were glossed over in the main body of the paper, which are useful for using the code:
Although the rewriting framework does not support dependently typed constants, we can automatically preprocess uses of eliminators like \mintinline{coq}{nat_rect} and \mintinline{coq}{list_rect} into nondependent versions.
The tactic that does this preprocessing is extensible via \Ltac{}'s reassignment feature.
Since pattern-matching compilation mixed with NbE requires knowing how many arguments a constant can be applied to, we must internally use a version of the recursion principle whose type arguments do not contain arrows; current preprocessing can handle recursion principles with either no arrows or one arrow in the motive.
Even though we will eventually plug in 0 for $v$, we jump through some extra hoops to ensure that our rewriter cannot cheat by rewriting away the ${}+0$ before reducing the recursion on $n$.

We can reduce this expression in three ways.

\subsubsection{Our Rewriter}
One lemma is required for rewriting with our rewriter:
\begin{minted}[fontsize=\small]{coq}
Lemma Z.add_0_r : forall z, z + 0 = z.
\end{minted}

Creating the rewriter takes about 12 seconds on the machine we used for running the performance experiments:
\begin{minted}[fontsize=\small]{coq}
Make myrew := Rewriter For (Z.add_0_r, eval_rect nat, eval_rect prod).
\end{minted}
Recall from \autoref{sec:explain-eval-rect} that \mintinline{coq}{eval_rect} is a definition provided by our framework for eagerly evaluating recursion associated with certain types.
It functions by triggering typeclass resolution for the lemmas reducing the recursion principle associated to the given type.
We provide instances for \texttt{nat}, \texttt{prod}, \texttt{list}, \texttt{option}, and \texttt{bool}.
Users may add more instances if they desire.

\subsubsection{\texorpdfstring{\texttt{setoid\_rewrite}}{setoid\_rewrite} and \texorpdfstring{\texttt{rewrite\_strat}}{rewrite\_strat}}
To give as many advantages as we can to the preexisting work on rewriting, we pre-reduce the recursion on \mintinline{coq}{nat}s using \taccbv{} before performing \texttt{setoid\_rewrite}.
(Note that \texttt{setoid\_rewrite} cannot itself perform reduction without generating large proof terms, and \texttt{rewrite\_strat} is not currently capable of sequencing reduction with rewriting internally due to bugs such as \coqbug{10923}.)
Rewriting itself is easy; we may use any of \texttt{repeat setoid\_rewrite Z.add\_0\_r}, \texttt{rewrite\_strat topdown Z.add\_0\_r}, or \texttt{rewrite\_strat bottomup Z.add\_0\_r}.

\subsection{Binders and Recursive Functions: LiftLetsMap}\label{sec:LiftLetsMap-more}\label{sec:micro:LiftLetsMap}

\begin{figure}
  {\small 
  $\begin{aligned}
    \text{map\_dbl}(\ell) & \defeq \begin{cases} [] & \text{if }\ell = [] \\
        \letin[{y := h + h}{}] & \text{if }\ell = h::t \\
        y :: \text{map\_dbl}(t) &
        \end{cases} \\
    \text{make}(n, m, v) & \defeq \begin{cases} [\underbrace{v, \ldots, v}_n] & \text{if }m = 0 \\
        \text{map\_dbl}(\text{make}(n, m-1, v)) & \text{if }m > 0
        \end{cases} \\
    \text{example}_{n, m} & \defeq \forall v,\ \text{make}(n, m, v) = []
  \end{aligned}$}%
  \caption{\label{fig:micro:LiftLetsMap:code}Initial code for binders and recursive functions}
\end{figure}

\begin{figure}
  \adjustbox{valign=t}{\resizebox{0.4\textwidth}{!}{\beginTikzpictureStamped[only marks]{
      \input{perf-UnderLetsPlus0.csv.md5} \space
  }
    \pgfplotsset{every axis legend/.append style={
        at={(0.5,-0.2)},
        anchor=north}}
    \begin{axis}[xlabel=$n\cdot m$,
        ylabel=time (s),
        scaled x ticks=false,
        ymax=27,
        xmax=12000,
        table/col sep=comma,
        table/x=param-0-nm]
      \addplot[mark=o,color=red]         table[y=rewrite-strat(bottomup-bottomup)-regression-exponential-user]{perf-LiftLetsMap.csv};
      \addplot[mark=triangle,color=red]  table[y=rewrite-strat(topdown-bottomup)-regression-exponential-user]{perf-LiftLetsMap.csv};
      \addplot[mark=square,color=red]    table[y=setoid-rewrite-regression-cubic-user]{perf-LiftLetsMap.csv};
      \addplot[mark=+,color=blue]        table[y=Rewrite-for-gen-user]{perf-LiftLetsMap.csv};
      \addplot[mark=x,color=ForestGreen] table[y=rewriting-user]{perf-LiftLetsMap.csv};
      \addplot[mark=*,color=purple]      table[y=cps+vm-compute-regression-quadratic-user]{perf-LiftLetsMap.csv};
      \legend{rewrite\_strat bottomup,rewrite\_strat topdown,repeat setoid\_rewrite,{Our approach including reification, cbv, etc.},Our approach (only rewriting),cps+vm\_compute}
    \end{axis}
  \end{tikzpicture}}}
  \caption{\label{fig:timing-LiftLetsMap}Benchmark with recursive functions}
\end{figure}

The next experiment uses the code in \autoref{fig:micro:LiftLetsMap:code}.
Note that the \letin{} binding blocks further reduction of map\_dbl when we iterate it $m$ times in make, and so we need to take care to preserve sharing when reducing here.

\autoref{fig:timing-LiftLetsMap} compares performance between our approach, \texttt{repeat setoid\_rewrite}, and two variants of \texttt{rewrite\_strat}.
Additionally, we consider another option, which was adopted by Fiat Cryptography at a larger scale: rewrite our functions to improve reduction behavior.
Specifically, both functions are rewritten in continuation-passing style, which makes them harder to read and reason about but allows standard VM-based reduction to achieve good performance.
The figure shows that \texttt{rewrite\_strat} variants are essentially unusable for this example, with \texttt{setoid\_rewrite} performing only marginally better, while our approach applied to the original, more readable definitions loses ground steadily to VM-based reduction on CPS'd code.
On the largest terms ($n \cdot m > 20,000$), the gap is 6s vs.\ 0.1s of compilation time, which should often be acceptable in return for simplified coding and proofs, plus the ability to mix proved rewrite rules with built-in reductions.
Note that about 99\% of the difference between the full time of our method and just the rewriting is spent in the final \taccbv{} at the end, used to denote our output term from reified syntax.
We blame this performance on the unfortunate fact that reduction in Coq is quadratic in the number of nested binders present; see Coq bug \coqbug{11151}.
This bug has since been fixed, as of Coq 8.14; see Coq PR~\coqpr{13537}.

We can perform this rewriting in four ways.

\subsubsection{Our Rewriter}
One lemma is required for rewriting with our rewriter:
\begin{minted}[fontsize=\small]{coq}
Lemma eval_repeat A x n
: @List.repeat A x ('n) = ident.eagerly nat_rect _ [] (λ k repeat_k, x :: repeat_k) ('n).
\end{minted}
Recall that the apostrophe marker (\verb|'|) is explained in \autoref{sec:explain-'}.
Recall again from \autoref{sec:explain-ident.eagerly} that we use \mintinline{coq}{ident.eagerly} to ask the reducer to simplify a case of primitive recursion by complete traversal of the designated argument's constructor tree.
Our current version only allows a limited, hard-coded set of eliminators with \mintinline{coq}{ident.eagerly} (\texttt{nat\_rect} on return types with either zero or one arrows, \texttt{list\_rect} on return types with either zero or one arrows, and \texttt{List.nth\_default}), but nothing in principle prevents automatic generation of the necessary code.

We construct our rewriter with
\begin{minted}[fontsize=\small]{coq}
Make myrew := Rewriter For (eval_repeat, eval_rect list, eval_rect nat)
  (with extra idents (Z.add)).
\end{minted}
On the machine we used for running all our performance experiments, this command takes about 13 seconds to run.
Note that all identifiers which appear in any goal to be rewritten must either appear in the type of one of the rewrite rules or in the tuple passed to \texttt{with extra idents}.

Rewriting is relatively simple, now.
Simply invoke the tactic \mintinline{coq}{Rewrite_for myrew}.
We support rewriting on only the left-hand-side and on only the right-hand-side using either the tactic \mintinline{coq}{Rewrite_lhs_for myrew} or else the tactic \mintinline{coq}{Rewrite_rhs_for myrew}, respectively.

\subsubsection{\texorpdfstring{\texttt{rewrite\_strat}}{rewrite\_strat}}

To reduce adequately using \texttt{rewrite\_strat}, we need the following two lemmas:
\begin{minted}[fontsize=\small]{coq}
Lemma lift_let_list_rect T A P N C (v : A) fls
: @list_rect T P N C (Let_In v fls) = Let_In v (fun v => @list_rect T P N C (fls v)).
Lemma lift_let_cons T A x (v : A) f
: @cons T x (Let_In v f) = Let_In v (fun v => @cons T x (f v)).
\end{minted}

Note that \mintinline{coq}{Let_In} is the constant we use for writing \letin{} expressions that do not reduce under $\zeta$.
Throughout most of this paper, anywhere that \letin{} appears, we have actually used \mintinline{coq}{Let_In} in the code.
It would alternatively be possible to extend the reification preprocessor to automatically convert \letin{} to \mintinline{coq}{Let_In}, but this may cause problems when converting the interpretation of the reified term with the prereified term, as Coq's conversion does not allow fine-tuning of when to inline or unfold \mintinline{coq}{let}s.

To rewrite, we start with \mintinline{coq}{cbv [example make map_dbl]} to expose the underlying term to rewriting.
One would hope that one could just add these two hints to a database \mintinline{coq}{db} and then write \texttt{rewrite\_strat (repeat (eval cbn [list\_rect]; try bottomup hints db))}, but unfortunately this does not work due to a number of bugs in Coq: \coqbug{10934}, \coqbug{10923}, \coqbug{4175}, \coqbug{10955}, and the potential to hit \coqbug{10972}.
Instead, we must put the two lemmas in separate databases, and then write \texttt{repeat (cbn [list\_rect]; (rewrite\_strat (try repeat bottomup hints db1)); (rewrite\_strat (try repeat bottomup hints db2)))}.
Note that the rewriting with \mintinline{coq}{lift_let_cons} can be done either top-down or bottom-up, but \texttt{rewrite\_strat} breaks if the rewriting with \mintinline{coq}{lift_let_list_rect} is done top-down.

\subsubsection{CPS and the VM}
If we want to use Coq's built-in VM reduction without our rewriter, to achieve the prior state-of-the-art performance, we can do so on this example, because it only involves partial reduction and not equational rewriting.
However, we must (a) module-opacify the constants which are not to be unfolded, and (b) rewrite all of our code in CPS.

Then we are looking at
\begin{align*}
    \text{map\_dbl\_cps}(\ell,k) & \defeq \begin{cases} k([]) & \text{if }\ell = [] \\
        \letin[{y := h +_\text{ax} h}{}] & \text{if }\ell = h::t \\
        \text{map\_dbl\_cps}(t, \\
        \qquad(\lambda ys, k(y :: ys)))
    \end{cases} \\
    \text{make\_cps}(n, m, v, k) & \defeq \begin{cases} k([\underbrace{v, \ldots, v}_n]) & \text{if }m = 0 \\
        \text{make\_cps}(n, m-1, v, & \text{if }m > 0 \\
        \quad(\lambda \ell,\text{map\_dbl\_cps}(\ell, k))
    \end{cases} \\
    \text{example\_cps}_{n, m} & \defeq \forall v,\ \text{make\_cps}(n, m, v, \lambda x.\,x) = []
\end{align*}

Then we can just run \tacvmcompute{}.
Note that this strategy, while quite fast, results in a stack overflow when $n \cdot m$ is larger than approximately $2.5\cdot 10^4$.
This is unsurprising, as we are generating quite large terms.
Our framework can handle terms of this size but stack-overflows on only slightly larger terms.

\subsubsection{Takeaway}

From this example, we conclude that \texttt{rewrite\_strat} is unsuitable for computations involving large terms with many binders, especially in cases where reduction and rewriting need to be interwoven, and that the many bugs in \texttt{rewrite\_strat} result in confusing gymnastics required for success.
The prior state of the art---writing code in CPS---suitably tweaked by using module opacity to allow \tacvmcompute{}, remains the best performer here, though the cost of rewriting everything is CPS may be prohibitive.
Our method soundly beats \texttt{rewrite\_strat}.
We are additionally bottlenecked on \taccbv{}, which is used to unfold the goal post-rewriting and costs about a minute on the largest of terms; see Coq bug \coqbug{11151} for a discussion on what is wrong with Coq's reduction here.

\subsection{SieveOfEratosthenes}\label{sec:Eratosthenes}\label{sec:micro:Eratosthenes}

The final experiment involves full reduction in computing the Sieve of Eratosthenes, taking inspiration on benchmark choice from \citet{Aehlig}.
We find in \autoref{fig:timing-SieveOfEratosthenes} that we are slower than \tacvmcompute{}, \tacnativecompute{}, and \taccbv{}, but faster than \mintinline{coq}{lazy}, and of course much faster than \tacsimpl{} and \taccbn{}, which are quite slow.

\begin{figure}
  \resizebox{0.4\textwidth}{!}{\beginTikzpictureStamped[only marks]{
      \input{perf-SieveOfEratosthenes.csv.md5} \space
  }
    \pgfplotsset{every axis legend/.append style={
        at={(0.5,-0.2)},
        anchor=north}}
    \begin{axis}[xlabel=upper bound,
        ylabel=time (s),
        ymax=65,
        xmax=5000,
        table/col sep=comma,
        table/x=param-n]
      \addplot[mark=square,color=red]       table[y=simpl-regression-cubic-user]                {perf-SieveOfEratosthenes.csv};
      \addplot[mark=triangle,color=red]     table[y=cbn-regression-cubic-user]                  {perf-SieveOfEratosthenes.csv};
      \addplot[mark=square*,color=red]      table[y=lazy-regression-quadratic-user]             {perf-SieveOfEratosthenes.csv};
      \addplot[mark=+,color=blue]           table[y=Rewrite-for-gen-user]                       {perf-SieveOfEratosthenes.csv};
      \addplot[mark=x,color=ForestGreen]    table[y=rewriting-user]                             {perf-SieveOfEratosthenes.csv};
      \addplot[mark=triangle*,color=orange] table[y=cbv-regression-quadratic-user]              {perf-SieveOfEratosthenes.csv};
      \addplot[mark=*,color=purple]         table[y=vm-compute-regression-quadratic-user]       {perf-SieveOfEratosthenes.csv};
      \addplot[mark=-,color=red]            table[y=native-compute(2)-regression-quadratic-real]{perf-SieveOfEratosthenes.csv};
      \legend{simpl,cbn,lazy,{Our approach including reification, cbv, etc.},Our approach (only rewriting),cbv,vm\_compute,native\_compute}
    \end{axis}
  \end{tikzpicture}}
  \caption{\label{fig:timing-SieveOfEratosthenes}Full evaluation, Sieve of Eratosthenes}
\end{figure}

We define the sieve using \mintinline{coq}{PositiveMap.t} and \mintinline{coq}{list Z}:
\begin{minted}[fontsize=\small]{coq}
Definition sieve' (fuel : nat) (max : Z) :=
 List.rev
  (fst
   (@nat_rect
    (λ _, list Z (* primes *) *
     PositiveSet.t (* composites *) *
     positive (* np (next_prime) *) ->
     list Z (* primes *) *
     PositiveSet.t (* composites *))
    (λ '(primes, composites, next_prime),
     (primes, composites))
    (λ _ rec '(primes, composites, np),
      rec
       (if (PositiveSet.mem np composites ||
            (Z.pos np >? max))%bool%Z
        then
         (primes, composites, Pos.succ np)
        else
         (Z.pos np :: primes,
          List.fold_right
           PositiveSet.add
           composites
           (List.map
            (λ n, Pos.mul (Pos.of_nat (S n)) np)
            (List.seq 0 (Z.to_nat(max/Z.pos np)))),
          Pos.succ np)))
    fuel
    (nil, PositiveSet.empty, 2%positive))).

Definition sieve (n : Z)
  := Eval cbv [sieve'] in sieve' (Z.to_nat n) n.
\end{minted}

We need four lemmas and an additional instance to create the rewriter:
\begin{minted}[fontsize=\small]{coq}
Lemma eval_fold_right A B f x ls :
@List.fold_right A B f x ls
= ident.eagerly list_rect _ _
    x
    (λ l ls fold_right_ls, f l fold_right_ls)
    ls.

Lemma eval_app A xs ys :
xs ++ ys
= ident.eagerly list_rect A _
    ys
    (λ x xs app_xs_ys, x :: app_xs_ys)
    xs.

Lemma eval_map A B f ls :
@List.map A B f ls
= ident.eagerly list_rect _ _
    []
    (λ l ls map_ls, f l :: map_ls)
    ls.

Lemma eval_rev A xs :
@List.rev A xs
= (@list_rect _ (fun _ => _))
    []
    (λ x xs rev_xs, rev_xs ++ [x])%list
    xs.

Scheme Equality for PositiveSet.tree.

Definition PositiveSet_t_beq
   : PositiveSet.t -> PositiveSet.t -> bool
  := tree_beq.

Global Instance PositiveSet_reflect_eqb
 : reflect_rel (@eq PositiveSet.t) PositiveSet_t_beq
 := reflect_of_brel
      internal_tree_dec_bl internal_tree_dec_lb.
\end{minted}

We then create the rewriter with
\begin{minted}[fontsize=\small]{coq}
Make myrew := Rewriter For
  (eval_rect nat, eval_rect prod, eval_fold_right,
   eval_map, do_again eval_rev, eval_rect bool,
   @fst_pair, eval_rect list, eval_app)
   (with extra idents (Z.eqb, orb, Z.gtb,
    PositiveSet.elements, @fst, @snd,
    PositiveSet.mem, Pos.succ, PositiveSet.add,
    List.fold_right, List.map, List.seq, Pos.mul,
    S, Pos.of_nat, Z.to_nat, Z.div, Z.pos, O,
    PositiveSet.empty))
  (with delta).
\end{minted}

To get \taccbn{} and \tacsimpl{} to unfold our term fully, we emit
\begin{minted}[fontsize=\small]{coq}
Global Arguments Pos.to_nat !_ / .
\end{minted}

\FloatBarrier\section{Fusing Compiler Passes}\label{sec:fusing-compiler-passes}

When we moved the
constant-folding rules
from before abstract interpretation to after it, the performance of our compiler on Word-by-Word Montgomery code synthesis decreased significantly.
(The generated code did not change.)
We discovered that the number of variable assignments in our intermediate code was quartic in the number of bits in the prime, while the number of variable assignments in the generated code is only quadratic.
The performance numbers we measured supported this theory: the overall running time of synthesizing code for a prime near $2^k$ jumped from $\Theta(k^2)$ to $\Theta(k^4)$ when we made this change.
We believe that fusing abstract interpretation with rewriting and partial evaluation would allow us to fix this asymptotic-complexity issue.

To make this situation more concrete, consider the following example:
Fiat Cryptography uses abstract interpretation to perform bounds analysis; each expression is associated with a range that describes the lower and upper bounds of values that expression might take on.
Abstract interpretation on addition works as follows: if we have that $x_\ell \le x \le x_u$ and $y_\ell \le y \le y_u$, then we have that $x_\ell + y_\ell \le x + y \le x_u + y_u$.
Performing bounds analysis on $+$ requires two additions.
We might have an arithmetic simplification that says that $x + y = x$ whenever we know that $0 \le y \le 0$.
If we perform the abstract interpretation and then the arithmetic simplification, we perform two additions (for the bounds analysis) and then two comparisons (to test the lower and upper bounds of $y$ for equality with 0).
We cannot perform the arithmetic simplification before abstract interpretation, because we will not know the bounds of $y$.
However, if we perform the arithmetic simplification for each expression after performing bounds analysis on its \emph{subexpressions} and only after this perform abstract interpretation on the resulting expression, then we need not use any additions to compute the bounds of $x + y$ when $0 \le y \le 0$, since the expression will just become $x$.

Another essential pass to fuse with rewriting and partial evaluation is let-lifting.
Unless all of the code is CPS-converted ahead of time, attempting to do let-lifting via rewriting, as must be done when using \mintinline{coq}{setoid_rewrite}, \mintinline{coq}{rewrite_strat}, or \Rtac, results in slower asymptotics.
This pattern is already apparent in the \texttt{LiftLetsMap} / ``Binders and Recursive Functions'' example in \autoref{sec:micro:LiftLetsMap}.
We achieve linear performance in $n\cdot m$ when ignoring the final \taccbv, while \mintinline{coq}{setoid_rewrite} and \mintinline{coq}{rewrite_strat} are both cubic.
The rewriter in \Rtac\space cannot possibly achieve better than $\mathcal{O}\left(n\cdot m^2\right)$ unless it can be sublinear in the number of rewrites, because our rewriter gets away with a constant number of rewrites (four), plus evaluating recursion principles for a total amount of work $\mathcal{O}(n\cdot m)$.
But without primitive support for let-lifting, it is instead necessary to lift the lets by rewrite rules, which requires
$\mathcal{O}\left(n\cdot m^2\right)$ rewrites just to lift the lets.
The analysis is thus: running \texttt{make} simply gives us $m$ nested applications of \texttt{map\_dbl} to a length-$n$ list.
To reduce a given call to \texttt{map\_dbl}, all existing let-binders must first be lifted (there are $n\cdot k$ of them on the $k$-innermost-call) across \texttt{map\_dbl}, one-at-a-time.
Then the \texttt{map\_dbl} adds another $n$ let binders, so we end up doing $\sum_{k=0}^{m} n\cdot k$ lifts, i.e., $n\cdot m(m+1)/2$ rewrites just to lift the lets.

\section{Experience vs.\ Lean and \texorpdfstring{\texttt{setoid\_rewrite}}{setoid\_rewrite}\label{sec:lean}}

Although all of our toy examples work with \texttt{setoid\_rewrite} or \texttt{rewrite\_strat} (until the terms get too big), even the smallest of examples in Fiat Cryptography fell over using these tactics.
When attempting to use \texttt{setoid\_rewrite} for partial evaluation and rewriting on unsaturated Solinas with 1 limb on small primes (such as $2^{61}-1$), we were able to get \texttt{setoid\_rewrite} to finish after about 100 seconds.
Trying to synthesize code for two limbs on slightly larger primes (such as $2^{107}-1$, which needs two limbs on a 64-bit machine) took about 10 minutes;
three limbs took just under 3.5 hours, and four limbs failed to synthesize with an out-of-memory error after using over 60 GB of RAM.
The widely used primes tend to have around five to ten limbs.
See \coqbug{13576} for more details and for updates.

The \texttt{rewrite\_strat} tactic, which does not require duplicating the entire goal at each rewriting step, fared a bit better.
Small primes with 1 limb took about 90 seconds, but further performance tuning of the typeclass instances dropped this time down to 11 seconds.
The bugs in \texttt{rewrite\_strat} made finding the right magic invocation quite painful, nonetheless; the invocation we settled on involved \emph{sixteen} consecutive calls to \texttt{rewrite\_strat} with varying arguments and strategies.
Two limbs took about 90 seconds, three limbs took a bit under 10 minutes, and four limbs took about 70 minutes and about 17 GB of RAM.
Extrapolating out the exponential asymptotics of the fastest-growing subcall to \texttt{rewrite\_strat} indicates that 5 limbs would take 11--12 hours, 6 limbs would take 10--11 days, 7 limbs would take 31--32 weeks, 8 limbs would take 13--14 years, 9 limbs would take 2--3 centuries, 10 limbs would take 6--7 millennia, and 15 limbs would take 2--3 times the age of the universe, and 17 limbs, the largest example we might find at present in the real world, would take over $1000\times$ the age of the universe!
See \coqbug{13708} for more details and updates.

This experiment using \verb|rewrite_strat| can be found online in the Coq source file at \texttt{src/fiat\_crypto\_via\_setoid\_rewrite\_standalone.v} on GitHub at \githublink{coq-community}{coq-performance-tests}.
To test with the two-limb prime $2^{107}-1$, change \verb|Goal goal| to \verb|Goal goal_of_size 2%nat| near the bottom of the file.

We also tried Lean, in the hopes that rewriting in Lean, specifically optimized for performance, would be up to the challenge.
Although Lean performed about 30\% better than Coq's \texttt{setoid\_rewrite} on the 1-limb example, taking a bit under a minute, it did not complete on the two-limb example even after four hours (after which we stopped trying), and a five-limb example was still going after 40 hours.

Our experiments with running \texttt{rewrite} in Lean on the Fiat Cryptography code can be found in the file \texttt{fiat-crypto-lean/src/fiat\_crypto.lean} on GitHub at \githublink{mit-plv}{fiat-crypto}[lean].
We used Lean version 3.4.2, commit cbd2b6686ddb, Release.
Run \texttt{make} in \texttt{fiat-crypto-lean} to run the one-limb example;
change \texttt{open ex} to \texttt{open ex2} to try the two-limb example, or to \texttt{open ex5} to try the five-limb example.

\section{Limitations and Preprocessing}\label{sec:implementation-and-usage}


We now note some details of the rewriting framework that were previously glossed over, which are useful for using the code or implementing something similar, but which do not add fundamental capabilities to the approach.
Although the rewriting framework does not support dependently typed constants, we can automatically preprocess uses of eliminators like \mintinline{coq}{nat_rect} and \mintinline{coq}{list_rect} into nondependent versions.
The tactic that does this preprocessing is extensible via \Ltac{}'s reassignment feature.
Since pattern-matching compilation mixed with NbE requires knowing how many arguments a constant can be applied to, internally we must use a version of the recursion principle whose type arguments do not contain arrows; current preprocessing can handle recursion principles with either no arrows or one arrow in motives.

Recall from \autoref{sec:explain-eval-rect} that \mintinline{coq}{eval_rect} is a definition provided by our framework for eagerly evaluating recursion associated with certain types.
It functions by triggering typeclass resolution for the lemmas reducing the recursion principle associated to the given type.
We provide instances for \texttt{nat}, \texttt{prod}, \texttt{list}, \texttt{option}, and \texttt{bool}.
Users may add more instances if they desire.

Recall again from \autoref{sec:explain-ident.eagerly} that we use \mintinline{coq}{ident.eagerly} to ask the reducer to simplify a case of primitive recursion by complete traversal of the designated argument's constructor tree.
Our current version only allows a limited, hard-coded set of eliminators with \mintinline{coq}{ident.eagerly} (\texttt{nat\_rect} on return types with either zero or one arrows, \texttt{list\_rect} on return types with either zero or one arrows, and \texttt{List.nth\_default}), but nothing in principle prevents automatic generation of the necessary code.

We define a constant \mintinline{coq}{Let_In} which we use for writing \letin{} expressions that do not reduce under $\zeta$ (Coq's reduction rule for \mintinline{coq}{let}-inlining).
Throughout most of this paper, anywhere that \letin{} appears, we have actually used \mintinline{coq}{Let_In} in the code.
It would alternatively be possible to extend the reification preprocessor to automatically convert \letin{} to \mintinline{coq}{Let_In}, but this strategy may cause problems when converting the interpretation of the reified term with the prereified term, as Coq's conversion does not allow fine-tuning of when to inline or unfold \mintinline{coq}{let}s.

\section{Reading the Code Supplement}\label{sec:CodeSupplement-more}

\catcode`\/=\active
\def/{\slash}%

We have attached both the code for implementing the rewriter, as well as a copy of Fiat Cryptography adapted to use the rewriting framework.
Both code supplements build with Coq versions 8.9--8.13, and they require that whichever OCaml was used to build Coq be installed on the system to permit building plugins.
(If Coq was installed via opam, then the correct version of OCaml will automatically be available.)
Both code bases can be built by running \texttt{make} in the top-level directory.

The performance data for both repositories are included at the top level as \texttt{.txt} and \texttt{.csv} files.

The performance data for the microbenchmarks can be rebuilt using \texttt{make perf-SuperFast perf-Fast perf-Medium} followed by \texttt{make perf-csv} to get the \texttt{.txt} and \texttt{.csv} files.
The microbenchmarks should run in about 24 hours when run with \texttt{-j5} on a 3.5 GHz machine.
There also exist targets \texttt{perf-Slow} and \texttt{perf-VerySlow}, but these take significantly longer.

The performance data for the macrobenchmark can be rebuilt from the Fiat Cryptography copy included by running \texttt{make perf -k}.
We ran this with \texttt{PERF\_MAX\_TIME=3600} to allow each benchmark to run for up to an hour; the default is 10 minutes per benchmark.
Expect the benchmarks to take over a week of time with an hour timeout and five cores.
Some tests are expected to fail, making \texttt{-k} a necessary flag.
Again, the \texttt{perf-csv} target will aggregate the logs and turn them into \texttt{.txt} and \texttt{.csv} files.

The entry point for the rewriter is the Coq source file \texttt{rewriter/src/Rewriter/Util/plugins/RewriterBuild.v}.

The rewrite rules used in Fiat Cryptography are defined in \texttt{fiat-crypto/src/Rewriter/Rules.v} and proven in \texttt{fiat-crypto/src/Rewriter/RulesProofs.v}.
Note that the Fiat Cryptography copy uses \verb|COQPATH| for dependency management, and \verb|.dir-locals.el| to set \verb|COQPATH| in emacs/PG; you must accept the setting when opening a file in the directory for interactive compilation to work.
Thus interactive editing either requires ProofGeneral or manual setting of \verb|COQPATH|.
The correct value of \verb|COQPATH| can be found by running \verb|make printenv|.

We will now go through this paper and describe where to find each reference in the code base.

\newcommand{\autocommanameref}[1]{\autoref{#1}, \nameref{#1}}

\subsection{Code from \autocommanameref{sec:intro}}

The P-384 curve is mentioned.
This is the curve with modulus $2^{384} - 2^{128} - 2^{96} + 2^{32} - 1$; its benchmarks can be found in files matching the glob \texttt{fiat-crypto/src/Rewriter/PerfTesting/Specific/generated/p2384m2128m296p232m1\_\_*\_\_word\_by\_word\_montgomery\_*}.
The output \texttt{.log} files are included in the tarball; the \texttt{.v} and \texttt{.sh} files are automatically generated in the course of running \texttt{make perf -k}.

\subsubsection{Code from \autocommanameref{sec:related-one}}

There is no code mentioned in this section.

\subsubsection{Code from \autocommanameref{sec:our-solution}}

We claimed that our solution meets five criteria.
We briefly justify each criterion with a sentence or a pointer to code:
\begin{itemize}
  \item
    We claimed that we \textbf{did not grow the trusted code base}.
    In any example file (of which a couple can be found in \texttt{rewriter/src/Rewriter/Rewriter/Examples/}), the \mintinline{coq}{Make} command creates a rewriter package.
    Running \texttt{Print Assumptions} on this new constant (often named \texttt{rewriter} or \texttt{myrew}) should demonstrate a lack of axioms.
    \texttt{Print Assumptions} may also be run on the proof that results from using the rewriter.
  \item
    We claimed \textbf{fast} partial evaluation with reasonable memory use; we assume that the performance graphs stand on their own to support this claim.
    Note that memory usage can be observed by making the benchmarks while passing \texttt{TIMED=1} to \texttt{make}.
  \item
    We claimed to allow reduction that \textbf{mixes} \emph{rules of the definitional equality} with \emph{equalities proven explicitly as theorems}; the ``rules of the definitional equality'' are, for example, $\beta$ reduction, and we assert that it should be self-evident that our rewriter supports this.
  \item
    We claimed to allow \textbf{rapid iteration} on rewrite rules with \emph{minimal verification overhead}.
    We invite the reader to alter the list of constants in any of the \mintinline{coq}{Make ... := Rewriter For ...} invocations in \texttt{rewriter/src/Rewriter/Rewriter/Examples/} or to alter the list of rewrite rules in \texttt{fiat-crypto/src/Rewriter/Rules.v} to experience iteration on rewrite rules.
  \item
    We claimed common-subterm \textbf{sharing preservation}.
    This is implemented by supporting the use of the \texttt{dlet} notation which is defined in \texttt{rewriter/src/Rewriter/Util/LetIn.v} via the \texttt{Let\_In} constant.
    We will come back to the infrastructure that supports this.
  \item
    We claimed \textbf{extraction of standalone partial evaluators}.
    The extraction is performed in the files \texttt{perf\_unsaturated\_solinas.v} and \texttt{perf\_word\_by\_word\_montgomery.v}, and the files \texttt{saturated\_solinas.v}, \texttt{unsaturated\_solinas.v}, and \texttt{word\_by\_word\_montgomery.v}, all in the directory \texttt{fiat-crypto/src/ExtractionOCaml/}.
    The OCaml code can be extracted and built using the target \texttt{make standalone-ocaml} (or \texttt{make perf-standalone} for the \texttt{perf\_} binaries).
    There may be some issues with building these binaries on Windows as some versions of \texttt{ocamlopt} on Windows seem not to support outputting binaries without the \texttt{.exe} extension.
\end{itemize}

We mention encoding pattern matching explicitly by adopting the performance-tuned approach of \citet{maranget2008compiling}; the code for this is in \texttt{rewriter/src/Rewriter/Rewriter/Rewriter.v} starting from the comment above \mintinline{coq}{Inductive decision_tree} and including the Gallina definitions \mintinline{coq}{eval_decision_tree} and \mintinline{coq}{compile_rewrites}.

We mention integration with abstract interpretation; the abstract-interpretation pass is implemented in \texttt{fiat-crypto/src/AbstractInterpretation/}; integration is achieved in rewrite rules in \texttt{fiat-crypto/src/Rewriter/Rules.v} making use of the various \mintinline{coq}{Local Notation}s defined in that file for \mintinline{coq}{ident.cast}.

We mention parametric higher-order abstract syntax (PHOAS); the definition of our datatype is \mintinline{coq}{Inductive expr} in module \mintinline{coq}{Compilers.expr} in \texttt{rewriter/src/Rewriter/Language/Language.v}.
We mention a let-lifting transformation threaded throughout reduction; this is \mintinline{coq}{Inductive UnderLets}, a monad defined in module \mintinline{coq}{Compilers.UnderLets} in the file \texttt{rewriter/src/Rewriter/Language/UnderLets.v}.

\subsection{Code from \autocommanameref{sec:motivating-example}}

The \texttt{prefixSums} example appears in the Coq source file \texttt{rewriter/src/Rewriter/Rewriter/Examples/PrefixSums.v}.
Note that we use \texttt{dlet} rather than \mintinline{coq}{let} in binding \texttt{acc'} so that we can preserve the \mintinline{coq}{let} binder even under $\iota$ reduction, which much of Coq's infrastructure performs eagerly.
Because we do not depend on the axiom of functional extensionality, we also in practice require \texttt{Proper} instances for each higher-order identifier saying that each constant respects function extensionality.
Although we glossed over this detail in the body of this paper, we also prove
\begin{minted}[fontsize=\small]{coq}
Global Instance: forall A B,
 Proper ((eq ==> eq ==> eq) ==> eq ==> eq ==> eq)
        (@fold_left A B).
\end{minted}

The \mintinline{coq}{Make} command is exposed in \texttt{rewriter/src/Rewriter/Util/plugins/RewriterBuild.v} and defined in \texttt{rewriter/src/Rewriter/Util/plugins/rewriter\_build\_plugin.mlg}.
Note that one must run \texttt{make} to create this latter file; it is copied over from a version-specific file at the beginning of the build.

\label{sec:code:eval-rect}%
\label{sec:code:ident.eagerly}%
The \verb|do_again|, \verb|eval_rect|, and \verb|ident.eagerly| constants are defined at the bottom of module \verb|RewriteRuleNotations| in \texttt{rewriter/src/Rewriter/Language/Pre.v}.

\subsection{Code from \autocommanameref{sec:structure}}


\subsubsection{Code from \autocommanameref{sec:nine-steps}}\label{sec:code:nine-steps}

We match the nine steps with functions from the source code:
\begin{enumerate}
  \item
    The given lemma statements are scraped for which named functions and types the rewriter package will support.
    This is performed by \texttt{rewriter\_scrape\_data} in the file \texttt{rewriter/src/Rewriter/Util/plugins/rewriter\_build.ml} which invokes the \Ltac{} tactic named \texttt{make\_scrape\_data} in a submodule in the source file \texttt{rewriter/src/Rewriter/Language/IdentifiersBasicGenerate.v} on a goal headed by the constant we provide under the name \texttt{Pre.ScrapedData.t\_with\_args} in \texttt{rewriter/src/Rewriter/Language/PreCommon.v}.
  \item
    Inductive types enumerating all available primitive types and functions are emitted.
    This step is performed by \texttt{rewriter\_emit\_inductives} in file \texttt{rewriter/src/Rewriter/Util/plugins/rewriter\_build.ml} invoking tactics, like \texttt{make\_base\_elim} in \texttt{rewriter/src/Rewriter/Language/IdentifiersBasicGenerate.v}, on goals headed by constants from \texttt{rewriter/src/Rewriter/Language/IdentifiersBasicLibrary.v}, including the constant \texttt{base\_elim\_with\_args} for example, to turn scraped data into eliminators for the inductives.
    The actual emitting of inductives is performed by code in the file \texttt{rewriter/src/Rewriter/Util/plugins/inductive\_from\_elim.ml}.
  \item
    Tactics generate all of the necessary definitions and prove all of the necessary lemmas for dealing with this particular set of inductive codes.
    This step is performed by the tactic \texttt{make\_rewriter\_of\_scraped\_and\_ind} in the source file \texttt{rewriter/src/Rewriter/Util/plugins/rewriter\_build.ml} which invokes the tactic \texttt{make\_rewriter\_all} defined in the file \texttt{rewriter/src/Rewriter/Rewriter/AllTactics.v} on a goal headed by the provided constant \texttt{VerifiedRewriter\_with\_ind\_args} defined in \texttt{rewriter/src/Rewriter/Rewriter/ProofsCommon.v}.
    The definitions emitted can be found by looking at the tactic \texttt{Build\_Rewriter} in \texttt{rewriter/src/Rewriter/Rewriter/AllTactics.v}, the \Ltac{} tactics \texttt{build\_package} in \texttt{rewriter/src/Rewriter/Language/IdentifiersBasicGenerate.v} and also in \texttt{rewriter/src/Rewriter/Language/IdentifiersGenerate.v} (there is a different tactic named \texttt{build\_package} in each of these files), and \texttt{prove\_package\_proofs\_via} which can be found in \texttt{rewriter/src/Rewriter/Language/IdentifiersGenerateProofs.v}.
  \item
    The statements of rewrite rules are reified and soundness and syntactic-well-formedness lemmas are proven about each of them.
    This is done as part of the previous step, when the tactic \texttt{make\_rewriter\_all} transitively calls \texttt{Build\_Rewriter} from \texttt{rewriter/src/Rewriter/Rewriter/AllTactics.v}.
    Reification is handled by the tactic \texttt{Build\_RewriterT} in \texttt{rewriter/src/Rewriter/Rewriter/Reify.v}, while soundness and the syntactic-well-formedness proofs are handled by the tactics \texttt{prove\_interp\_good} and \texttt{prove\_good} respectively, both in the source file \texttt{rewriter/src/Rewriter/Rewriter/ProofsCommonTactics.v}.
  \item
    The definitions needed to perform reification and rewriting and the lemmas needed to prove correctness are assembled into a single package that can be passed by name to the rewriting tactic.
    This step is also performed by \texttt{make\_rewriter\_of\_scraped\_and\_ind} in the source file \texttt{rewriter/src/Rewriter/Util/plugins/rewriter\_build.ml}.
\end{enumerate}

When we want to rewrite with a rewriter package in a goal, the following steps are performed, with code in the following places:
\begin{enumerate}
  \item
    We rearrange the goal into a closed logical formula: all free-variable quantification in the proof context is replaced by changing the equality goal into an equality between two functions (taking the free variables as inputs).
    Note that it is not actually an equality between two functions but rather an \texttt{equiv} between two functions, where \texttt{equiv} is a custom relation we define indexed over type codes that is equality up to function extensionality.
    This step is performed by the tactic \texttt{generalize\_hyps\_for\_rewriting} in \texttt{rewriter/src/Rewriter/Rewriter/AllTactics.v}.
  \item
    We reify the side of the goal we want to simplify, using the inductive codes in the specified package.
    That side of the goal is then replaced with a call to a denotation function on the reified version.
    This step is performed by the tactic \texttt{do\_reify\_rhs\_with} in \texttt{rewriter/src/Rewriter/Rewriter/AllTactics.v}.
  \item
    We use a theorem stating that rewriting preserves denotations of well-formed terms to replace the denotation subterm with the denotation of the rewriter applied to the same reified term.
    We use Coq's built-in full reduction (\tacvmcompute{}) to reduce the application of the rewriter to the reified term.
    This step is performed by the tactic \texttt{do\_rewrite\_with} in \texttt{rewriter/src/Rewriter/Rewriter/AllTactics.v}.
  \item
    Finally, we run \taccbv{} (a standard call-by-value reducer) to simplify away the invocation of the denotation function on the concrete syntax tree from rewriting.
    This step is performed by the tactic \texttt{do\_final\_cbv} in \texttt{rewriter/src/Rewriter/Rewriter/AllTactics.v}.
\end{enumerate}
These steps are put together in the tactic \texttt{Rewrite\_for\_gen} in \texttt{rewriter/src/Rewriter/Rewriter/AllTactics.v}.

The expression language $e$ corresponds to the inductive \texttt{expr} type defined in the module \texttt{Compilers.expr} in \texttt{rewriter/src/Rewriter/Language/Language.v}.

\subsubsection*{Our Approach in More Than Nine Steps}

As the nine steps of \autoref{sec:nine-steps} do not exactly match the code, we describe here a more accurate version of what is going on.
For ease of readability, we do not clutter this description with references to the code supplement, instead allowing the reader to match up the steps here with the more coarse-grained ones in \autoref{sec:nine-steps} or \autoref{sec:code:nine-steps}.

In order to allow easy invocation of our rewriter, a great deal of code (about 6500 lines) needed to be written.
Some of this code is about reifying rewrite rules into a form that the rewriter can deal with them in.
Other code is about proving that the reified rewrite rules preserve interpretation and are well-formed.
We wrote some plugin code to automatically generate the inductive type of base-type codes and identifier codes, as well as the two variants of the identifier-code inductive used internally in the rewriter.
One interesting bit of code that resulted was a plugin that can emit an inductive declaration given the Church encoding (or eliminator) of the inductive type to be defined.
We wrote a great deal of tactic code to prove basic properties about these inductive types, from the fact that one can unify two identifier codes and extract constraints on their type variables from this unification, to the fact that type codes have decidable equality.
Additional plugin code was written to invoke the tactics that construct these definitions and prove these properties, so that we could generate an entire rewriter from a single command, rather than having the user separately invoke multiple commands in sequence.

In order to build the precomputed rewriter, the following actions are performed:
\begin{enumerate}
    \item
    The terms and types to be supported by the rewriter package are scraped from the given lemmas.
    \item
    An inductive type of codes for the types is emitted, and then three different versions of inductive codes for the identifiers are emitted (one with type arguments, one with type arguments supporting pattern type variables, and one without any type arguments, to be used internally in pattern-matching compilation).
    \item
    Tactics generate all of the necessary definitions and prove all of the necessary lemmas for dealing with this particular set of inductive codes.
    Definitions cover categories like ``Boolean equality on type codes'' and ``how to extract the pattern type variables from a given identifier code,'' and lemma categories include ``type codes have decidable equality'' and ``the types being coded for have decidable equality'' and ``the identifiers all respect function extensionality.''
    \item
    The rewrite rules are reified, and we prove interpretation-correctness and well-formedness lemmas about each of them.
    \item
    The definitions needed to perform reification and rewriting and the lemmas needed to prove correctness are assembled into a single package that can be passed by name to the rewriting tactic.
    \item
    The denotation functions for type and identifier codes are marked for early expansion in the kernel via the \mintinline{coq}{Strategy} command;
    this is necessary for conversion at \mintinline{coq}{Qed}-time to perform reasonably on enormous goals.
\end{enumerate}

When we want to rewrite with a rewriter package in a goal, the following steps are performed:
\begin{enumerate}
    \item
    We use \mintinline{coq}{etransitivity} to allow rewriting separately on the left- and right-hand-sides of an equality.
    Note that we do not currently support rewriting in non-equality goals, but this is easily worked around using \texttt{let v := open\_constr:(\_) in replace <some term> with v} and then rewriting in the second goal.
    \item
    We revert all hypotheses mentioned in the goal, and change the form of the goal from a universally quantified statement about equality into a statement that two functions are extensionally equal.
    Note that this step will fail if any hypotheses are functions not known to respect function extensionality via typeclass search.
    \item
    We reify the side of the goal that is not an existential variable using the inductive codes in the specified package; the resulting goal equates the denotation of the newly reified term with the original evar.
    \item
    We use a lemma stating that rewriting preserves denotations of well-formed terms to replace the goal with the rewriter applied to our reified term.
    We use \tacvmcompute{} to prove the well-formedness side condition reflectively.
    We use \tacvmcompute{} again to reduce the application of the rewriter to the reified term.
    \item
    Finally, we run \taccbv{} to unfold the denotation function, and we instantiate the evar with the resulting rewritten term.
\end{enumerate}

There are a couple of steps that contribute to the trusted code base.
We must trust that the rewriter package we generate from the rewrite rules in fact matches the rewrite rules we want to rewrite with.
This involves partially trusting the scraper, the reifier, and the glue code.
We must also trust the VM we use for reduction at various points in rewriting.
Otherwise, everything is checked by Coq.

\subsubsection{Code from \autocommanameref{sec:pattern-matching-compilation-and-evaluation}}

The pattern-matching compilation step is done by the tactic \texttt{CompileRewrites} in \texttt{rewriter/src/Rewriter/Rewriter/Rewriter.v}, which just invokes the Gallina definition named \texttt{compile\_rewrites} with ever-increasing amounts of fuel until it succeeds.
(It should never fail for reasons other than insufficient fuel, unless there is a bug in the code.)
The workhorse function here is \texttt{compile\_rewrites\_step}.

The decision-tree evaluation step is done by the definition \texttt{eval\_rewrite\_rules}, also in the file \texttt{rewriter/src/Rewriter/Rewriter/Rewriter.v}.
The correctness lemmas are the theorem \texttt{eval\_rewrite\_rules\_correct} in the file \texttt{rewriter/src/Rewriter/Rewriter/InterpProofs.v} and the theorem \texttt{wf\_eval\_rewrite\_rules} in \texttt{rewriter/src/Rewriter/Rewriter/Wf.v}.
Note that the second of these lemmas, not mentioned in the paper, is effectively saying that for two related syntax trees, \texttt{eval\_rewrite\_rules} picks the same rewrite rule for both.
(We actually prove a slightly weaker lemma, which is a bit harder to state in English.)

The third step of rewriting with a given rule is performed by the definition \texttt{rewrite\_with\_rule} in \texttt{rewriter/src/Rewriter/Rewriter/Rewriter.v}.
The correctness proof goes by the name \texttt{interp\_rewrite\_with\_rule} in \texttt{rewriter/src/Rewriter/Rewriter/InterpProofs.v}.
Note that the well-formedness-preservation proof for this definition in inlined into the proof of the lemma \verb|wf_eval_rewrite_rules| mentioned above.

The inductive description of decision trees is \verb|decision_tree| in \texttt{rewriter/src/Rewriter/Rewriter/Rewriter.v}.

The pattern language is defined as the inductive \verb|pattern| in \texttt{rewriter/src/Rewriter/Rewriter/Rewriter.v}.
Note that we have a \verb|Raw| version and a typed version; the pattern-matching compilation and decision-tree evaluation of \citet{Aehlig} is an algorithm on untyped patterns and untyped terms.
We found that trying to maintain typing constraints led to headaches with dependent types.
Therefore when doing the actual decision-tree evaluation, we wrap all of our expressions in the dynamically typed \verb|rawexpr| type and all of our patterns in the dynamically typed \verb|Raw.pattern| type.
We also emit separate inductives of identifier codes for each of the \verb|expr|, \verb|pattern|, and \verb|Raw.pattern| type families.

We partially evaluate the partial evaluator defined by \verb|eval_rewrite_rules| in the \Ltac{} tactic \verb|make_rewrite_head| in \texttt{rewriter/src/Rewriter/Rewriter/Reify.v}.

\subsubsection{Code from \autocommanameref{sec:thunk-eval-subst-term}}

The type $\text{NbE}_t$ mentioned in this paper is not actually used in the code; the version we have is described in \autoref{sec:under-lets} as the definition \verb|value'| in \texttt{rewriter/src/Rewriter/Rewriter/Rewriter.v}.

The functions \verb|reify| and \verb|reflect| are defined in \texttt{rewriter/src/Rewriter/Rewriter/Rewriter.v} and share names with the functions in the paper.
The function \texttt{reduce} is named \verb|rewrite_bottomup| in the code, and the closest match to NbE is \verb|rewrite|.

\subsection{Code from \autocommanameref{sec:scaling}}

\subsubsection{Code from \autocommanameref{sec:PHOAS}}

The inductives \verb|type|, \verb|base_type| (actually the inductive type \verb|base.type.type| in the supplemental code), and \verb|expr|, as well as the definition \verb|Expr|, are all defined in \texttt{rewriter/src/Rewriter/Language/Language.v}.
The definition \verb|denoteT| is the fixpoint \verb|type.interp| (the fixpoint \verb|interp| in the module \verb|type|) in \texttt{rewriter/src/Rewriter/Language/Language.v}.
The definition \verb|denoteE| is \verb|expr.interp|, and \verb|DenoteE| is the fixpoint \verb|expr.Interp|.

As mentioned above, \verb|nbeT| does not actually exist as stated but is close to \verb|value'| in \texttt{rewriter/src/Rewriter/Rewriter/Rewriter.v}.
The functions \verb|reify| and \verb|reflect| are defined in \texttt{rewriter/src/Rewriter/Rewriter/Rewriter.v} and share names with the functions in the paper.
The actual code is somewhat more complicated than the version presented in the paper, due to needing to deal with converting well-typed-by-construction expressions to dynamically typed expressions for use in decision-tree evaluation and also due to the need to support early partial evaluation against a concrete decision tree.
Thus the version of \verb|reflect| that actually invokes rewriting at base types is a separate definition \verb|assemble_identifier_rewriters|, while \verb|reify| invokes a version of \verb|reflect| (named \verb|reflect|) that does not call rewriting.
The function named \texttt{reduce} is what we call \verb|rewrite_bottomup| in the code; the name \verb|Rewrite| is shared between this paper and the code.
Note that we eventually instantiate the argument \verb|rewrite_head| of \verb|rewrite_bottomup| with a partially evaluated version of the definition named \verb|assemble_identifier_rewriters|.
Note also that we use fuel to support \verb|do_again|, and this is used in the definition \verb|repeat_rewrite| that calls \verb|rewrite_bottomup|.

The correctness proofs are \verb|InterpRewrite| in the Coq source file \texttt{rewriter/src/Rewriter/Rewriter/InterpProofs.v} and \verb|Wf_Rewrite| in \texttt{rewriter/src/Rewriter/Rewriter/Wf.v}.

Packages containing rewriters and their correctness theorems are in the record \verb|VerifiedRewriter| in \texttt{rewriter/src/Rewriter/Rewriter/ProofsCommon.v};
a package of this type is then passed to the tactic \verb|Rewrite_for_gen| from \texttt{rewriter/src/Rewriter/Rewriter/AllTactics.v} to perform the actual rewriting.
The correspondence of the code to the various steps in rewriting is described in the second list of \autoref{sec:code:nine-steps}.

\subsubsection{Code from \autocommanameref{sec:under-lets}}

To run the P-256 example in the copy of Fiat Cryptography attached as a code supplement, after building the library, run the code
\begin{minted}[fontsize=\small]{coq}
Require Import Crypto.Rewriter.PerfTesting.Core.
Require Import Crypto.Util.Option.

Import WordByWordMontgomery.
Import Core.RuntimeDefinitions.

Definition p : params
  := Eval compute in invert_Some (of_string "2^256-2^224+2^192+2^96-1" 64).

Goal True.
  (* Successful run: *)
  Time let v := (eval cbv
    -[Let_In
      runtime_nth_default
      runtime_add runtime_sub runtime_mul runtime_opp runtime_div runtime_modulo
      RT_Z.add_get_carry_full RT_Z.add_with_get_carry_full RT_Z.mul_split]
    in (GallinaDefOf p)) in
    idtac.
  (* Unsuccessful OOM run: *)
  Time let v := (eval cbv
    -[(*Let_In*)
      runtime_nth_default
      runtime_add runtime_sub runtime_mul runtime_opp runtime_div runtime_modulo
      RT_Z.add_get_carry_full RT_Z.add_with_get_carry_full RT_Z.mul_split]
    in (GallinaDefOf p)) in
    idtac.
Abort.
\end{minted}

The \verb|UnderLets| monad is defined in the file \texttt{rewriter/src/Rewriter/Language/UnderLets.v}.

The definitions \verb|nbeT'|, \verb|nbeT|, and \verb|nbeT_with_lets| are in \texttt{rewriter/src/Rewriter/Rewriter/Rewriter.v} and are named \verb|value'|, \verb|value|, and \verb|value_with_lets|, respectively.

\subsubsection{Code from \autocommanameref{sec:side-conditions}}

The ``variant of pattern variable that only matches constants'' is actually special support for the reification of \verb|ident.literal| (defined in the module \verb|RewriteRuleNotations| in \texttt{rewriter/src/Rewriter/Language/Pre.v}) threaded throughout the rewriter.
The apostrophe notation \verb|'| is also introduced in the module \verb|RewriteRuleNotations| in \texttt{rewriter/src/Rewriter/Language/Pre.v}.
The support for side conditions is handled by permitting rewrite-rule-replacement expressions to return \verb|option expr| instead of \verb|expr|, allowing the function \verb|expr_to_pattern_and_replacement| in the file \texttt{rewriter/src/Rewriter/Rewriter/Reify.v} to fold the side conditions into a choice of whether to return \verb|Some| or \verb|None|.

\subsubsection{Code from \autocommanameref{sec:abs-int}}

The abstract-interpretation pass is defined in \texttt{fiat-crypto/src/AbstractInterpretation/}, and the rewrite rules handling abstract-interpretation results are the Gallina definitions \verb|arith_with_casts_rewrite_rulesT|, as well as \verb|strip_literal_casts_rewrite_rulesT|, as well as \verb|fancy_with_casts_rewrite_rulesT|, and finally as well as \verb|mul_split_rewrite_rulesT|, all defined in \texttt{fiat-crypto/src/Rewriter/Rules.v}.

The \verb|clip| function is the definition \verb|ident.cast| in \texttt{fiat-crypto/src/Language/PreExtra.v}.

\subsection{Code from \autocommanameref{sec:evaluation}}

\subsubsection{Code from \autocommanameref{sec:iteration}}
The old continuation-passing-style versions of verified arithmetic functions can be found in the folder \texttt{fiat-crypto/src/ArithmeticCPS/}, while the new versions can be found in the folder \texttt{fiat-crypto/src/Arithmetic/}.

The rewrite rules
for reassociating arithmetic
can be found in \mintinline{coq}{arith_rewrite_rulesT} starting at the comment ``We reassociate some multiplication of small constants'' in \texttt{fiat-crypto/src/Rewriter/Rules.v}.

The
following
frontend constructs
are in \verb|all_ident_named_interped| defined in \texttt{fiat-crypto/src/Language/IdentifierParameters.v}.
\begin{itemize}
\item
  The multiplication primitives are \mintinline{coq}{with_name ident_Z_mul_split Z.mul_split} as well as \mintinline{coq}{with_name ident_Z_mul_high Z.mul_high}, as well as the various Coq expressions \mintinline{coq}{with_name ident_fancy_mulXX ident.fancy.mulXX} for each \mintinline{coq}{X} being either \texttt{l} or \texttt{h}.
\item
  The ``comment'' function is both \mintinline{coq}{with_name ident_comment (@ident.comment)} as well as \mintinline{coq}{with_name ident_comment_no_keep (@ident.comment_no_keep)}.
\item
  The bitwise exclusive-or is \mintinline{coq}{with_name ident_Z_lxor Z.lxor}.
\item
  The special identity function which prints in the backend as a call to some inline assembly is \mintinline{coq}{with_name ident_value_barrier (@Z.value_barrier)}.
\end{itemize}

The rules about bitmasking operations
can be found in \mintinline{coq}{arith_with_casts_rewrite_rulesT} in \texttt{fiat-crypto/src/Rewriter/Rules.v} and involve \mintinline{coq}{Z.land} and \mintinline{coq}{Z.lor}.

The compiler configuration about conditional-move instructions is the flag \texttt{--cmovznz-by-mul} defined in \texttt{fiat-crypto/src/CLI.v}.
The \texttt{if}-statement using the thus-defined \mintinline{coq}{use_mul_for_cmovznz} is in \texttt{src/PushButtonSynthesis/Primitives.v}.

The rewrite rules for the new backends
are defined by \mintinline{coq}{fancy_with_casts_rewrite_rulesT} and \mintinline{coq}{mul_split_rewrite_rulesT} as well as \mintinline{coq}{multiret_split_rewrite_rulesT} as well as \mintinline{coq}{noselect_rewrite_rulesT} in \texttt{fiat-crypto/src/Rewriter/Rules.v}.
The special function \mintinline{coq}{Z.combine_at_bitwidth} is defined in \texttt{fiat-crypto/src/Util/ZUtil/Definitions.v}.
The designation of \mintinline{coq}{Z.combine_at_bitwidth} as an identifier that should be inlined occurs by listing it in the definition \mintinline{coq}{var_like_idents} in the source file \texttt{fiat-crypto/src/Language/IdentifierParameters.v}.

The rules involving carries mentioned in
\autocommanameref{sec:fusing-compiler-passes} are in \mintinline{coq}{arith_with_casts_rewrite_rulesT} in \texttt{fiat-crypto/src/Rewriter/Rules.v}.

\subsubsection{Code from \autocommanameref{sec:micro}}

This code is found in the files in \texttt{rewriter/src/Rewriter/Rewriter/Examples/}.
We ran the microbenchmarks using the code in \texttt{rewriter/src/Rewriter/Rewriter/Examples/PerfTesting/Harness.v} together with some \texttt{Makefile} cleverness.

The code for \autoref{fig:timing-Plus0Tree} from \autocommanameref{sec:micro:Plus0Tree} can be found in \texttt{Plus0Tree.v}.

The code for \autoref{fig:timing-UnderLetsPlus0} from \autocommanameref{sec:micro:UnderLetsPlus0} can be found in \texttt{UnderLetsPlus0.v}.


The code for \autoref{fig:timing-LiftLetsMap} from \autocommanameref{sec:micro:LiftLetsMap} can be found in \texttt{LiftLetsMap.v}.

The code for \autoref{fig:timing-SieveOfEratosthenes} from \autocommanameref{sec:micro:Eratosthenes} can be found in \texttt{SieveOfEratosthenes.v}.

\subsubsection{Code from \autocommanameref{sec:macro}}

The rewrite rules are defined in \texttt{fiat-crypto/src/Rewriter/Rules.v} and proven in the file \texttt{fiat-crypto/src/Rewriter/RulesProofs.v}.
They are turned into rewriters in the various files in \texttt{fiat-crypto/src/Rewriter/Passes/}.
The shared inductives and definitions are defined in the Coq source file \texttt{fiat-crypto/src/Language/IdentifiersBasicGENERATED.v}, the Coq source file \texttt{fiat-crypto/src/Language/IdentifiersGENERATED.v}, and finally also the Coq source file \texttt{fiat-crypto/src/Language/IdentifiersGENERATEDProofs.v}.
Note that we invoke the subtactics of the \mintinline{coq}{Make} command manually to increase parallelism in the build and to allow a shared language across multiple rewriter packages.

\subsection{Code from \autocommanameref{sec:implementation-and-usage}}\label{sec:code-from-implementation-and-usage}

The \Ltac{} hooks for extending the preprocessing of eliminators are \mintinline{coq}{reify_preprocess_extra} and \mintinline{coq}{reify_ident_preprocess_extra} in a submodule of \texttt{rewriter/src/Rewriter/Language/PreCommon.v}.
These hooks are called by \mintinline{coq}{reify_preprocess} and \mintinline{coq}{reify_ident_preprocess} in a submodule of \texttt{rewriter/src/Rewriter/Language/Language.v}.
Some recursion lemmas for use with these tactics are defined in the \verb|Thunked| module in \texttt{fiat-crypto/src/Language/PreExtra.v}.
These tactics are overridden in the file \texttt{fiat-crypto/src/Language/IdentifierParameters.v}.

The typeclass associated to \mintinline{coq}{eval_rect} ({c.f.} \autoref{sec:code:eval-rect}) is \mintinline{coq}{rules_proofs_for_eager_type} defined in \texttt{rewriter/src/Rewriter/Language/Pre.v}.
The instances we provide by default are defined in a submodule of \texttt{src/Rewriter/Language/PreLemmas.v}.

The hard-coding of the eliminators for use with \mintinline{coq}{ident.eagerly} ({c.f.} \autoref{sec:code:ident.eagerly}) is done in the tactics \mintinline{coq}{reify_ident_preprocess} and \mintinline{coq}{rewrite_interp_eager} in \texttt{rewriter/src/Rewriter/Language/Language.v}, in the inductive type \mintinline{coq}{restricted_ident} and the typeclass \mintinline{coq}{BuildEagerIdentT} in \texttt{rewriter/src/Rewriter/Language/Language.v}, and in the \Ltac{} tactic with the name of \mintinline{coq}{handle_reified_rewrite_rules_interp} defined in the file \texttt{rewriter/src/Rewriter/Rewriter/ProofsCommonTactics.v}.

The \mintinline{coq}{Let_In} constant is defined in \texttt{rewriter/src/Rewriter/Util/LetIn.v}.

\end{document}